\documentclass[aps,prb,amsmath,amssymb,footinbib,showpacs,twocolumn,superscriptaddress]{revtex4-2}
\usepackage{amsmath}
\usepackage{amssymb}
\usepackage{amsthm}
\usepackage{setspace}
\usepackage{graphicx}
\usepackage{braket}
\usepackage{mathrsfs}
\usepackage{xcolor}
\usepackage{float}
\usepackage{ulem}
\definecolor{ao(english)}{rgb}{0.0, 0.5, 0.0}
\definecolor{mygreen}{rgb}{0.29, 0.5, 0.1}
\usepackage[colorlinks = true,linkcolor = mygreen,urlcolor  = blue,citecolor = blue,anchorcolor = blue]{hyperref}
\usepackage[utf8]{inputenc}
\usepackage[english]{babel}
\usepackage{bm}
\usepackage{csquotes}
\MakeOuterQuote{"}
\usepackage{placeins}
\usepackage{comment}
\usepackage{multirow}
\usepackage{booktabs}

\begin{document}

\title{Majorana modes in graphene strips: polarization, wavefunctions, disorder, and Andreev states}
\author{Shubhanshu Karoliya}
\affiliation{School of Physical Sciences, Indian Institute of Technology Mandi, Mandi 175005, India.}
\author{Sumanta Tewari}
\affiliation{Department of Physics and Astronomy, Clemson University, Clemson, South Carolina 29634, USA.}
\author{Gargee Sharma}
\affiliation{Department of Physics, Indian Institute of Technology Delhi, Hauz Khas, New Delhi 110016, India.}

\begin{abstract}
Topologically protected Majorana zero modes (MZMs) have attracted intense interest due to their potential application in fault-tolerant quantum computation (TQC). Graphene nanoribbons, with tunable edge terminations and compatibility with planar device architectures, offer a promising alternative to semiconductor nanowires. Here we present a comprehensive theoretical study of finite graphene strips with armchair, zigzag, and nearly square geometries, proximitized by an s-wave superconductor and subject to Rashba spin–orbit coupling, Zeeman fields, and disorder. Using exact diagonalization of the Bogoliubov-de Gennes tight-binding Hamiltonian, we analyze Majorana polarization, \textcolor{black}{a Pfaffian-based periodic-disorder topological indicator (PDI)}, low-energy spectra, and real-space wavefunctions to identify the non-trivial topological phases supporting MZMs and distinguish them from partially separated Andreev bound states (psABS) or the quasi-Majoranas. We systematically chart the robustness of these modes across geometries and disorder regimes, finding that armchair strips with short zigzag edges provide the most stable platform. Our results unify polarization \textcolor{black}{and PDI} diagnostics with spatial wavefunction analysis and disorder effects, yielding concrete design guidelines for graphene-based topological superconductors.
\end{abstract}

\maketitle

\section{Introduction}

Majorana fermions (MFs), proposed as real solutions to the Dirac equation~\cite{Majorana1937}, remain elusive as fundamental particles despite ongoing high-energy physics searches~\cite{das2023search}. In condensed matter, however, MFs can arise as zero-energy quasiparticles in certain superconductors, where they exhibit non-Abelian exchange statistics~\cite{moore1991nonabelions,read2000paired,nayak19962n}, and offer a potential route towards fault-tolerant topological quantum computation (TQC)~\cite{Kitaev2003,Nayak2008}. The canonical platform for realizing such states employs a spin-orbit coupled semiconductor proximitized by an $s$-wave superconductor under a Zeeman field~\cite{sau2010generic,sau2010non,oreg2010helical,lutchyn2010majorana}, which supports topological Majorana bound states (MBSs) localized at the boundaries of the system. The Rashba spin-orbit coupled one-dimensional nanowire is the most widely studied realization of this proposal. In the topological regime, MBSs are predicted to generate a quantized zero-bias conductance peak of $2e^2/h$ in tunneling spectroscopy~\cite{sengupta2001midgap,law2009majorana,flensberg2010tunneling}, robust to variations in magnetic field and gate voltage. While numerous experiments have reported features consistent with this picture~\cite{mourik2012signatures,deng2012anomalous,das2012zero,rokhinson2012fractional,churchill2013superconductor,finck2013anomalous,deng2016majorana,zhang2017ballistic,chen2017experimental,nichele2017scaling,albrecht2017transport,o2018hybridization,shen2018parity,sherman2017normal,vaitiekenas2018selective,albrecht2016exponential,Yu_2021,microsoft2025interferometric,mondal2025distinguishing,glodzik2020measure,rossi2020majorana,tanaka2024theory,tanaka2011symmetry,sharma2016tunneling,tanaka2009manipulation}, similar zero-bias signatures can also emerge from topologically trivial states~\cite{Brouwer2012,Mi2014,Bagrets2012,pikulin2012zero,ramon2012transport,pan2020physical,moore2018two,moore2018quantized,vuik2018reproducing,Stanescu_Robust,ramon_Jorge2106exceptional,ramon2019nonhermitian,Jorge2019supercurrent,sharma2020hybridization,zeng2020feasibility,zeng2022partially,ramon2020from,Jorge2021distinguishing,zhang2021,DasSarma2021,Frolov2021}, such as partially separated Andreev bound states (psABSs)~\cite{moore2018two,moore2018quantized} stabilized by disorder. This ambiguity has motivated the search for complementary diagnostics and alternative material platforms.

Graphene has emerged as a particularly attractive candidate. While pristine graphene is gapless and exhibits negligible intrinsic spin-orbit coupling, engineering finite-width graphene nanoribbons (GNRs) in proximity to an $s$-wave superconductor can open a superconducting gap and enable topological phases when Rashba spin-orbit coupling (RSOC) and a Zeeman field are introduced. Such graphene–superconductor hybrids combine experimental tunability with planar scalability, and several works have analyzed Majorana bound states in graphene nanoribbons and flakes~\cite{ma2025graphene,kaladzhyan2017formation,kaladzhyan2017majorana,PhysRevX.5.041042, laubscher2020majorana,wang2018strain, manesco2019effective}. In parallel, Majorana polarization (MP) has been proposed as a powerful diagnostic tool for distinguishing MZMs from trivial modes~\cite{sticlet2012spin,sedlmayr2015visualizing,sedlmayr2016majorana,karoliya2025majorana}. Yet, most studies have focused either on polarization diagnostics in idealized systems or on graphene wavefunctions in the clean limit, without a unified analysis of polarization diagnostics, disorder robustness, and possible competing Andreev-bound-state physics.
\textcolor{black}{At the same time, recent developments have emphasized the importance of complementing local Majorana diagnostics with global topological invariants, particularly in finite and disordered superconducting systems where trivial near-zero-energy states may mimic Majorana signatures. In this context, Pfaffian-based periodic disorder invariants (PDIs), constructed through boundary-twist formulations of periodically repeated disordered supercells, provide a powerful real-space approach for characterizing class-D topological superconductivity even in the absence of translational symmetry~\cite{eissele2025topological,roy2026pfaffian}.} 
%Motivated by these developments, in the present work we combine Majorana polarization, spectral diagnostics, real-space wavefunction analysis, and Pfaffian-based topology to investigate the emergence and robustness of Majorana-supporting regimes in proximitized graphene strips.}

There remains a critical gap: a comprehensive, self-consistent treatment that combines (i) Majorana polarization, \textcolor{black}{(ii) PDI}, (iii) real-space wavefunction morphology, (iv) disorder effects, and (v) analysis of psABS (quasi-Majoranas) and Andreev bound states, all within the graphene geometry context, is largely missing from the literature. This is especially important because graphene’s two-dimensional nature, edge types (zigzag, armchair), and richer bandstructure nuances imply that simplistic extrapolations from nanowires may fail.

In this work we address this gap. We perform a systematic investigation of finite-size graphene strips with different edge terminations: armchair, zigzag, and nearly square geometries, proximitized by an s-wave superconductor, subject to Rashba spin-orbit coupling and both in-plane and out-of-plane magnetic fields. Using exact diagonalization of the BdG Hamiltonian, we compute phase diagrams of MP, track bulk gap closings/reopenings, and image real-space localization for low-energy modes in clean and disordered regimes. By cross-correlating MP with wavefunction localization and monitoring psABS proliferation, we establish practical criteria for distinguishing genuine MZMs from trivial near-zero modes. Our analysis shows that the combination of edge geometry, magnetic field orientation, and disorder strength plays a decisive role in accessing and stabilizing topological superconducting phases supporting Majorana zero modes.  We shows that armchair strips with short zigzag ends best support sharply localized, well-separated MZMs that remain stable under moderate disorder; by contrast, zigzag and square geometries display narrower windows and are more easily compromised by disorder-induced psABSs. Our analysis provides not only a consistency check across multiple diagnostics but also design principles for scalable graphene-based platforms.

The remainder of this paper is organized as follows. In Sec. II, we introduce the tight-binding model for a graphene-based system with proximity-induced superconductivity, mapped onto an effective $p$-wave superconductor. \textcolor{black}{Section III reviews the concept of Majorana polarization, including both the chiral-symmetry-based formulation~\cite{sticlet2012spin} and the more general definition applicable to systems in which chiral symmetry is absent~\cite{sedlmayr2015visualizing}, \textcolor{black}{together with a Pfaffian-based periodic-disorder indicator, obtained from the corresponding periodically repeated disordered supercell, used as a global diagnostic of the finite disordered graphene strips.}~\cite{eissele2025topological,roy2026pfaffian}.} Section IV describes the finite-size graphene geometries considered in this work. In Sec. V, we present our results for Majorana zero modes in the three geometries, analyzing the role of magnetic field orientation, the importance of edge termination, and disorder strength. Finally, Sec. VI summarizes our conclusions.  

%------------------------------------------------------------------------------------------------------------------------

%------------------------------------------------------------------------------------------------------------------------------------------------------
\section{Model}

\textcolor{black}{We model graphene strips with proximity-induced superconductivity using the following tight-binding Hamiltonian~\cite{kaladzhyan2017formation}. The Hamiltonian in Eq.~(1) is defined on the underlying graphene honeycomb (hexagonal) lattice, consisting of two interpenetrating triangular sublattices, conventionally denoted A and B (see schematic in Fig.~\ref{f0}). The finite armchair, zigzag, and nearly square strip geometries studied in this work are obtained by imposing the corresponding boundary terminations on this parent honeycomb lattice:}
\begin{equation} \label{e1}
    H_0 = \sum_j \Psi^{\dagger}_j \big[V(\mathbf{x}_j) \tau^z - \mu \tau^z - \Delta \tau^x \big] \Psi_j - t \sum_{\langle i,j \rangle} \Psi^{\dagger}_i \tau^z \Psi_j
\end{equation}
The above Hamiltonian is written in a four-component Nambu basis, $\Psi_j = {(c_{j,\uparrow},c_{j,\downarrow},c_{j,\downarrow}^{\dagger},-c_{j,\uparrow}^{\dagger})}^T$. Here $c^\dagger_{j,\sigma}$ creates an electron with spin $\sigma$ at the site $j$. We use Pauli matrices, $\vec{\tau} =(\tau_x,\tau_y,\tau_z)$ for the particle-hole subspace, and $\vec{\sigma}=(\sigma_x,\sigma_y,\sigma_z)$ for the spin subspace. %We write complete tight-binding Hamiltonian in combination with three terms that are $H_0$, $H_b$, and, $H_R$ where 
In the above equation, $\tau^k \equiv \tau_k \otimes \sigma_0$ where $k = x, y, z$; $t$ denotes the hopping strength, $\mu$ represents the chemical potential, $V(\mathbf{x}_j)$ represents the position dependent potential, and $\Delta$ is the induced superconducting pairing potential. The summation $\langle i,j\rangle$ is restricted to only the nearest neighbors. Unless otherwise specified, throughout this manuscript, all the energy scales are expressed in terms of the hopping strength $t$, and all the distances are expressed in terms of the nearest-neighbor distance $a$.

The next term in the Hamiltonian describes the effect of the Zeeman magnetic field $\vec{h}$, expressed as:
\begin{equation} \label{e2}
    H_B = \sum_j \Psi_j^{\dagger} \vec{h} \cdot \vec{\sigma} \Psi_j,
\end{equation}
where $\vec{h} \cdot \vec{\sigma} \equiv h_k \tau_0 \otimes \sigma_k$ for $k = x, y, z$. {Throughout this work we retain only the Zeeman coupling of the magnetic
field and neglect its orbital effect. For in-plane magnetic fields, this approximation is well justified, since an external field parallel to the graphene sheet produces negligible magnetic flux through the monolayer and therefore induces no significant orbital coupling. In contrast, a perpendicular magnetic field would generate Landau quantization and strong orbital effects; hence, in the out-of-plane case, the Zeeman term is understood to arise from magnetic proximity to a ferromagnetic insulator rather than from an externally applied field. Such an exchange-induced Zeeman field acts only on the electron spin and does not generate a vector potential in graphene, thereby allowing us to include the Zeeman splitting without introducing orbital couplings in the tight-binding model.} The third term, $H_R$, accounts for the effective Rashba spin-orbit interaction of strength $\alpha$, and is given by:
\begin{equation} \label{e3}
    H_R = i \alpha \sum_{<i,j>} \Psi^{\dagger}_i (\vec{\delta_{ij}} \times \vec{\sigma}) \cdot \hat{z} \tau^z \Psi_j
\end{equation}
\textcolor{black}{Here, $\vec{\delta}_{ij}=\mathbf{r}_j-\mathbf{r}_i$ denotes the nearest-neighbor bond vector connecting lattice sites $i$ and $j$. For the graphene honeycomb lattice, each site has three such nearest-neighbor bond vectors linking opposite sublattices (A and B), and these vectors determine the direction-dependent Rashba spin--orbit coupling in Eq.~\ref{e3}}.
The complete Hamiltonian is thus expressed as:
\begin{equation} \label{e3.5} 
    H = H_0 + H_R + H_B.
\end{equation}
In our calculations, unless otherwise specified, we fix $\alpha = 0.5t$, $\Delta = 0.2t$. 
\textcolor{black}{In pristine graphene, the intrinsic atomic spin-orbit coupling is extremely weak. In the present work, the Rashba spin-orbit coupling parameter should therefore be understood as an effective interaction generated through interface engineering rather than the intrinsic SOC of isolated graphene. Experimentally, sizable SOC has been induced in graphene through proximity coupling to materials with strong spin-orbit interaction, such as transition-metal dichalcogenides and heavy-metal substrates, which generate enhanced Rashba-type terms through interfacial hybridization and inversion-symmetry breaking~\cite{wang2015strong,wang2016origin,marchenko2012giant}. Additional enhancement mechanisms include adatom decoration and chemical functionalization~\cite{weeks2011engineering}. Accordingly, the SOC strength employed in our model represents an experimentally motivated effective parameter for engineered proximitized graphene nanoribbon platforms.}
$V(\mathbf{x}_i)$ is the effective position-dependent disorder potential, which is modeled by considering \textcolor{black}{$\mathcal{Z}$} randomly distributed short-range impurities. The potential profile of an isolated impurity located at position $\mathbf{x}_i$ is 
\begin{equation}
V_\mathrm{imp}^{(i)}(\mathbf{x}) = A_i \exp\left(-\frac{|\mathbf{x}-\mathbf{x}_i|}{\lambda}\right), \label{Eq:Vimp}
\end{equation}
where $\lambda$ is the characteristic decay length of the impurity potential and $A_i$ is a random amplitude characterized by a Gaussian probability distribution.  The net disorder potential at position $\mathbf{x}$ is then given by 
\begin{equation}
V(\mathbf{x}) = V_0 \sum_{i=1}^{\textcolor{black}{\mathcal{Z}}}A_i\exp\left(-\frac{|\mathbf{x}-\mathbf{x}_i|}{\lambda}\right),  \label{Eq:Vdis}
\end{equation}
where $V_0$ allows us to tune the overall magnitude of the disorder for a given disorder configuration. 

\section{Diagnostics of Majorana modes and topology}

\subsection{Majorana polarization in graphene strips}
Before we proceed to discuss the geometry-dependent behavior of Majorana modes in graphene nanoribbons, we discuss the definitions of Majorana polarization that are applicable to Graphene. 

Graphene has a bipartite lattice structure composed of A and B sublattices. In the following Nambu basis:
\[
(c^\dagger_{iA\uparrow}, c^\dagger_{iB\uparrow}, c^\dagger_{iA\downarrow}, c^\dagger_{iB\downarrow}, c_{iA\downarrow}, c_{iB\downarrow}, -c_{iA\uparrow}, -c_{iB\uparrow}),
\]where \( i \) labels the lattice site, let us suppose that the solution of the BdG Hamiltonian at energy \( \epsilon \) is
\[
\psi_i(\epsilon) = \left(u_{iA\uparrow}, u_{iB\uparrow}, u_{iA\downarrow}, u_{iB\downarrow}, v_{iA\downarrow}, v_{iB\downarrow}, v_{iA\uparrow}, v_{iB\uparrow}\right),
\]
where \( u_{i\alpha} \) and \( v_{i\alpha} \) are particle and hole components for the $\alpha$ sublattice. Due to particle-hole symmetry, a corresponding eigenstate exists at \( -\epsilon \), given by
\[
\psi_i(-\epsilon) = \left(v_{iA\uparrow}^*, v_{iB\uparrow}^*, v_{iA\downarrow}^*, v_{iB\downarrow}^*, u_{iA\downarrow}^*, u_{iB\downarrow}^*, u_{iA\uparrow}^*, u_{iB\uparrow}^*\right).
\]
We define a Majorana basis as:
\begin{align}
\gamma_{i\alpha\sigma}^1 &= \frac{1}{\sqrt{2}} \left(c^\dagger_{i\alpha\sigma} + c_{i\alpha\sigma} \right), \nonumber\\
\gamma_{i\alpha\sigma}^2 &= \frac{i}{\sqrt{2}} \left(c^\dagger_{i\alpha\sigma} - c_{i\alpha\sigma} \right), \label{Eq:MajoranaBasis}
\end{align}
where $\alpha$ is the sublattice index. The Majorana polarization on the A sublattice at site \( i \) is defined as the difference in probabilities for the two Majorana components \( \gamma^1_{iA\sigma} \) and \( \gamma^2_{iA\sigma} \). Analogously, for the B sublattice at site \( j \), we compute the difference between \( \gamma^1_{iB\sigma} \) and \( \gamma^2_{iB\sigma} \). The rationale behind this is that the sum of probabilities corresponds to a full fermionic state, while the difference corresponds to a Majorana state.
Using the above basis, the polarization becomes~\cite{sticlet2012spin}:
\[
\mathcal{P}_i^\alpha = 2 \, \mathrm{Re} \left[ u_{i\alpha\downarrow} v_{i\alpha\downarrow}^* - u_{i\alpha\uparrow} v_{i\alpha\uparrow}^* \right].
\]
The local Majorana polarization at a given site on sublattice $\alpha$ and energy $\omega$ is defined as:
\begin{align}
    \mathcal{P}^\alpha_i(\omega) &= 2\sum_k \delta(\omega - \epsilon_k) \, \mathrm{Re}\left[u^k_{i\alpha\downarrow} v^{k*}_{i\alpha\downarrow} - u^k_{i\alpha\uparrow} v^{k*}_{i\alpha\uparrow} \right],
\end{align}
where the summation runs over all eigenstates of the Bogoliubov-de Gennes (BdG) Hamiltonian, and $u^k_{i\alpha\sigma}$, $v^k_{i\alpha\sigma}$ are the particle and hole components of the $k$-th eigenstate at site $i$ with sublattice $\alpha$, and spin $\sigma$. 
One may then sum over a suitable region $\mathcal{R}$ and calculate the total polarization in the region. 
This definition of MP works well if the Hamiltonian preserves chiral symmetry, i.e., the A and B sublattices \textit{do not see} each other. If chiral symmetry is broken, for example by a transverse Rashba term~\cite{tewari2012topological}, then the Majorana modes localized at A and B sublattices may couple to one other and thus it is useful to employ a general definition of Majorana polarization. 

In a recent work, Sedlmayr \textit{et al.}~\cite{sedlmayr2015visualizing} introduced a general definition of Majorana polarization valid in systems with or without chiral symmetry. The Bogoliubov-de Gennes Hamiltonian ($H_\mathrm{BdG}$), which describes excitations of a superconductor, inherently has built-in particle-hole symmetry. This symmetry operation is represented by a particle-hole operator $\mathcal{C}$ that anticommutes with the $H_\mathrm{BdG}$. Since a true Majorana bound state, $|\psi\rangle$, is an equal superposition of both particle and hole components, it is an eigenstate of both the Hamiltonian $H_\mathrm{BdG}$ and the particle-hole operator $\mathcal{C}$, and therefore must occur at exactly zero energy, i.e., ($[H_\mathrm{BdG},\mathcal{C}]_{\pm}|\psi\rangle=0$, where $\pm$ refer to commutator/anti-commutator). 
The particle-hole operator the takes the form $\mathcal{C} = \nu_0\sigma_y\tau_y \mathcal{K}$, where $\sigma_y$ and $\tau_y$ are Pauli-$y$ matrices for spin and electron-hole degree of freedoms, $\nu_0$ acts on the sublattice degree of freedom, and $\mathcal{K}$ represents complex conjugation. The local Majorana polarization $\mathcal{P}(\mathbf{r}_i,\epsilon)$ at a particular site $\mathbf{r}_i$ is defined to be the expectation value of the particle hole operator, i.e., $\mathcal{P}(\mathbf{r}_i,\epsilon_j)=\langle\psi(\mathbf{r}_i,\epsilon_j) | \mathcal{C}|\psi(\mathbf{r}_i,\epsilon_j)\rangle$. Furthermore, if an MBS is localized in a region $\mathcal{R}$ in space, the normalized polarization should be a complex number with unit magnitude, i.e., 
\begin{align}
\mathcal{P}_\mathcal{R}(\epsilon_j)=\frac{\sum_{i\in\mathcal{R}}\langle\psi(\mathbf{r}_i,\epsilon_j) | \mathcal{C}|\psi(\mathbf{r}_i,\epsilon_j)\rangle}{\sum_{i\in\mathcal{R}}\langle\psi(\mathbf{r}_i,\epsilon_j) | \psi(\mathbf{r}_i,\epsilon_j)\rangle} = e^{i\xi}.
\label{Eq:Pregion1}
\end{align}
To evaluate the polarization at a finite frequency $\omega$ we sum over the eigenstates weighted by a Dirac-delta function, i.e., 
\begin{align}
    \mathcal{P}_\mathcal{R}(\omega)=\sum_j\left[\frac{\sum_{i\in\mathcal{R}}\langle\psi(\mathbf{r}_i,\epsilon_j) | \mathcal{C}|\psi(\mathbf{r}_i,\epsilon_j)\rangle }{\sum_{i\in\mathcal{R}}\langle\psi(\mathbf{r}_i,\epsilon_j) | \psi(\mathbf{r}_i,\epsilon_j)\rangle }\right] \cdot \delta(\epsilon_j-\omega)
    \label{Eq:P_region2}
\end{align}
Here, the summation $j$ is over all the eigenstates, and the Dirac-Delta function is implemented by a narrow Gaussian factor of width $\sigma$. 
Note that $\mathcal{P}_\mathcal{R}$ is a complex number, with real and imaginary parts, $\mathcal{P}_\mathcal{R}=\mathcal{P}_\mathcal{R}^x+i\mathcal{P}_\mathcal{R}^y$. Reference~\cite{awoga2024identifying} recently introduced the quantity $\mathcal{P}_l^* \mathcal{P}_r$ as a measure for identifying non-local correlations between the two halves of a system (\textcolor{black}{left- and right-half}). While $\mathcal{P}_l$ and $\mathcal{P}_r$ are individually complex, their conjugate product, $P = \mathcal{P}_l^* \mathcal{P}_r = \mathcal{P}_l \mathcal{P}_r^*$, was found to be real in that work. In our earlier study of one-dimensional nanowires and quasi-1D systems~\cite{karoliya2025majorana}, we also observed $P$ to be real. However, in our recent investigation of a finite-size graphene strip platform for Majorana modes, we find that $\mathcal{P}$ is generally complex. Consequently, we use $|\mathcal{P}|$ as a diagnostic tool for Majorana mode identification,~\cite{sedlmayr2015visualizing}.
\textcolor{black}{Specifically, we define $\mathcal{P}_{l}$, $\mathcal{P}_{r}$, $\mathcal{P}_{u}$, and $\mathcal{P}_{d}$ as the Majorana polarizations evaluated over the left, right, upper, and lower halves of the finite graphene strip, respectively, obtained by choosing the corresponding spatial region $\mathcal R$ in Eq.~(\ref{Eq:P_region2}). The strip may be viewed as a rectangular finite system that can be partitioned equally in two independent ways. Division by a vertical in-plane axis passing through the center yields the left and right regions, corresponding to $\mathcal{P}_{l}$ and $\mathcal{P}_{r}$. Division by a horizontal in-plane axis through the center yields the upper and lower regions, corresponding to $\mathcal{P}_{u}$ and $\mathcal{P}_{d}$. Throughout the manuscript, the notations $\mathcal{P}^{\nu}$ and $\mathcal{P}_{\nu}$, with $\nu=u,d,l,r$, are used interchangeably to denote the same regional polarization.}

\textcolor{black}{Another important remark about the polarization is that the magnitude of the energy-resolved Majorana polarization is not a universally bounded quantity, since its numerical evaluation depends on the finite-width broadening used to represent the Dirac-delta function in Eq.~\ref{Eq:P_region2}. In the present work, the delta function is approximated by a narrow Gaussian of width $\sigma$ centered at $\omega=0$, such that low-energy states within an energy window set by $\sigma$ contribute to the summed polarization. Consequently, increasing $\sigma$ or increasing the density of near-zero-energy states generally enhances the absolute magnitude of the plotted polarization. The polarization amplitude therefore should not be interpreted as a normalized topological invariant. Instead, the physically relevant information lies in the relative evolution, enhancement, suppression, and spatial structure of the polarization across parameter space, together with its consistency with spectral gap behavior and real-space wavefunction localization. For the fixed broadening used throughout this work, the maximum polarization values observed are of order $\sim 25$.}

{Before analyzing the geometry and topological properties of the finite-size graphene nanoribbon, it is useful to clarify whether the fermion-doubling constraints of lattice Dirac systems apply to our model. The fermion-doubling problem is a generic artifact of lattice regularizations of massless chiral fermions, under very general assumptions of locality, Hermiticity, and discrete translational invariance \cite{nielsen1981absence1, nielsen1981absence2, nielsen1981nogotheorem, beenakker2023tangent}. \textcolor{black}{The Hamiltonian used in this work is not obtained by naively discretizing a single continuum Dirac Hamiltonian. Instead, we work
directly with the honeycomb-lattice tight-binding model of graphene. Therefore, the fermion-doubling problem associated with lattice regularizations of an isolated continuum Dirac fermion does not arise as a numerical artifact in our calculation.
However, in pristine graphene, this doubling is built in as the well-known valley degeneracy, i.e., the honeycomb lattice supports two inequivalent Dirac cones at the $\mathbf{K}$ and $\mathbf{K}$' points, which are related by lattice symmetries and carry opposite Berry-phase winding, so that the low-energy theory consists of two species of massless Dirac fermions. In the proximitized graphene nanoribbon considered here, however, the relevant low-energy degrees of freedom are
not independent continuum Dirac cones, but the discrete subbands of the
full finite-width tight-binding BdG Hamiltonian.} Specifically, the finite-size graphene strip imposes boundary conditions that explicitly admix $\mathbf{K}$ and $\mathbf{K}$' valley components of the wavefunction, lifting valley as a good quantum number and producing a single ladder of transverse 1D subbands rather than two independent valley copies \cite{brey2006electronic}. Secondly, the Rashba spin-orbit coupling, a uniform Zeeman field, and proximity-induced s-wave pairing lift the residual spin degeneracy so that, for suitable values of chemical potential ($\mu/t$) and Zeeman energy ($h/t$), only one helical subband crosses the Fermi level. 

The presence of disorder further breaks translational invariance along the ribbon, so the conditions of the Nielsen-Ninomiya theorem are not satisfied, and no additional low-energy doublers are enforced. }

\begin{figure*}
    \centering
    \includegraphics[width=1.79\columnwidth]{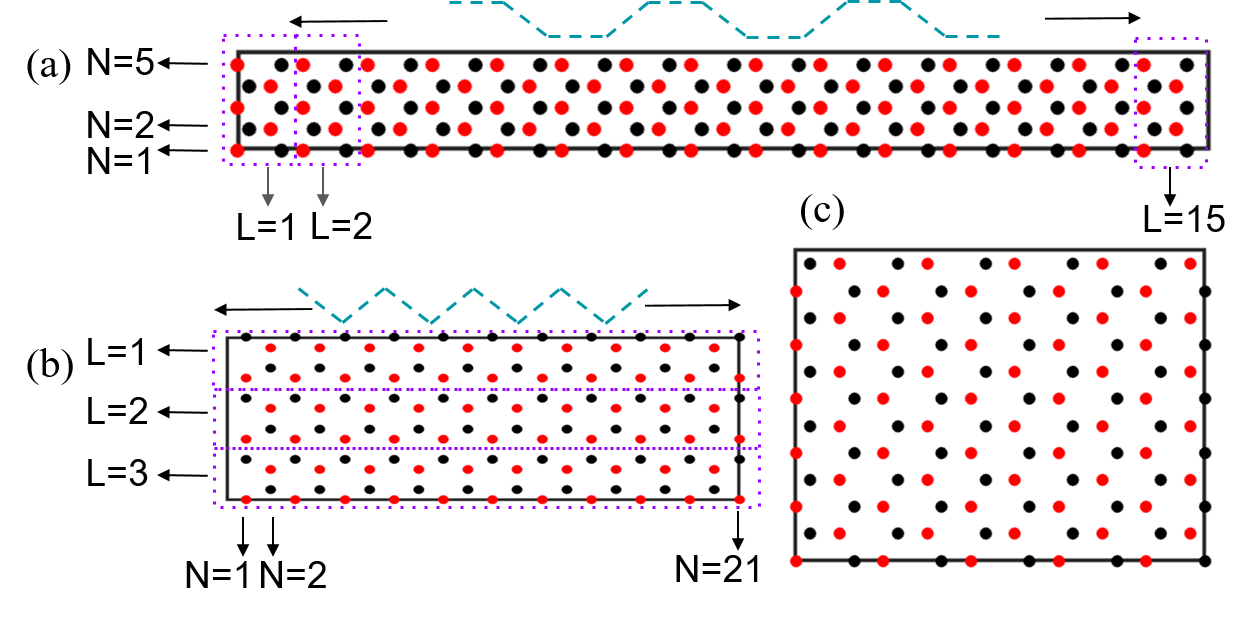}
    \caption{\textcolor{black}{Finite-size graphene strips constructed from the underlying honeycomb (hexagonal) lattice with two sublattices: A (red) and B (black).} (a) Armchair type strip (A-strip) with N = 5 and L = 15 (zigzag short edges). (b) Zigzag type strip (Z-type) with N = 21 and L = 3 (armchair short edges). (c) Nearly square type strip (S-type) with N = 12 and L = 5. N represents the number of layers, and L represents the number of unit cells in a system. Red and black dots denote the sublattices. \textcolor{black}{The strips are oriented such that the armchair edge lies along the x direction.}}
    \label{f0}
\end{figure*}
\begin{figure*}
    \centering
    \includegraphics[width=1.99\columnwidth]{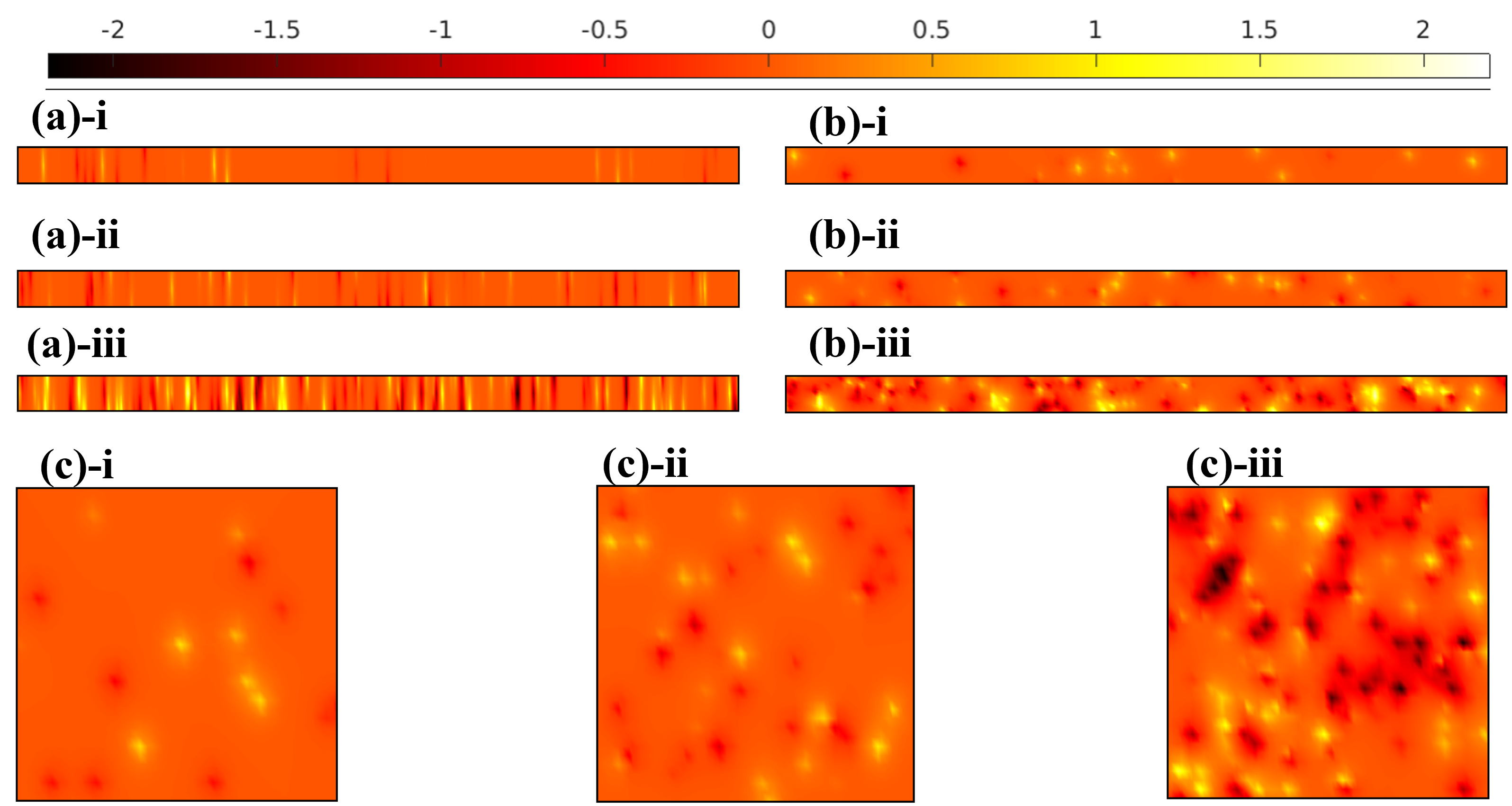}
    \caption{Spatial profiles of disorder potentials used in our calculations for different ribbon geometries: (a) armchair strip, (b) zigzag strip, and (c) square strip. Panels (i), (ii), and (iii) correspond to increasing disorder strengths: weak disorder with $\mathcal{Z} = 20$ impurities and $V_0 = 1.0$ (top row), moderate disorder with $\mathcal{Z} = 50$ impurities and $V_0 = 1.0$ (middle row), and strong disorder with $\mathcal{Z} = 150$ impurities and $V_0 = 1.5$ (bottom row). The color scale represents the disorder potential values, ranging from $-2.2$ to $+2.2$.}
    \label{fig:potential_profile}
\end{figure*}

\subsection{Pfaffian periodic-disorder invariant}

\textcolor{black}{To complement the Majorana polarization and real-space wave-function analysis, we evaluate a Pfaffian-based topological invariant for the periodically repeated disordered graphene strip~\cite{eissele2025topological,roy2026pfaffian}. We emphasize that Majorana polarization is a local, wave-function-based diagnostic, whereas the Pfaffian invariant \textcolor{black}{of the periodically repeated supercell} is a global topological diagnostic of the Bogoliubov-de Gennes (BdG) Hamiltonian. The combined use of these quantities is useful because finite disordered systems can host near-zero-energy states with sizable Majorana polarization even when those states are not protected by a topological bulk gap.
The BdG Hamiltonian introduced in Eq.~\ref{e1} is written in the Nambu basis
$\Psi_j=\left(c_{j\uparrow}, c_{j\downarrow}, c^\dagger_{j\downarrow}, -c^\dagger_{j\uparrow} \right)^T$
and satisfies the intrinsic particle-hole symmetry
$\mathcal{C}H_{\rm BdG}\mathcal{C}^{-1} = -H_{\rm BdG}$,
where
\begin{equation}
\mathcal{C} = U_{\mathcal C}\mathcal{K} = \nu_0\sigma_y\tau_y\mathcal{K}.
\end{equation}
For a class-D BdG Hamiltonian, the Pfaffian invariant can be defined at particle-hole invariant boundary twists. We construct
\begin{equation}
B(\phi)=H(\phi)U_{\mathcal C},
\end{equation}
where $\phi$ is the boundary twist. At $\phi=0$ and $\phi=\pi$, particle-hole symmetry implies
\begin{equation}
B^T(\phi)=-B(\phi),
\end{equation}
so that the Pfaffian $\mathrm{Pf}[B(\phi)]$ is well defined. A change in the sign of the Pfaffian product can occur only when the corresponding quasiparticle gap closes, and therefore signals a topological phase transition.
For a clean translationally invariant one-dimensional superconductor, the same invariant is usually written in momentum space as
\begin{equation}
\mathcal{Q}= {\rm sgn} \left[ {\rm Pf}\,B(k=0)\, {\rm Pf}\,B(k=\pi) \right],
\end{equation}
where $k=0$ and $k=\pi$ are the particle-hole symmetric momenta. In the present disordered finite strip, momentum is not a good quantum number. We therefore use the periodic-disorder construction: a finite disordered segment is treated as a supercell and periodically repeated. Equivalently, one introduces a twist phase across the boundary connecting the last and first unit cells~\cite{roy2026pfaffian}.
}

\textcolor{black}{
It is important to emphasize that the Majorana polarization is evaluated directly for the finite open-boundary graphene strip, whereas the Pfaffian invariant is defined for the periodically repeated disordered supercell obtained through the boundary-twist construction described above. In this sense, the periodicity entering the Pfaffian calculation is imposed at the supercell level: a finite disordered graphene segment is treated as a repeating unit, thereby restoring an effective translational structure suitable for defining the class-D $\mathbb{Z}_2$ invariant. Consequently, the Pfaffian invariant should be interpreted as the bulk topological invariant of the periodically repeated disordered graphene superlattice rather than of the isolated finite strip alone.} 
\textcolor{black}{For the zigzag-strip geometry, implementing this periodic closure additionally imposes a geometric constraint on the transverse width. Since the short edges of the zigzag strip are of armchair type, periodic boundary conditions require the boundary hopping to reconnect opposite graphene sublattices ($A\leftrightarrow B$). This occurs naturally only for even values of the transverse layer number $N$. For odd $N$, the periodic closure would instead connect identical sublattices ($A\leftrightarrow A$ and $B\leftrightarrow B$), which is incompatible with the nearest-neighbor hopping structure of the graphene honeycomb lattice. For this reason, we use an even transverse width ($N=170$) in the zigzag-strip calculations involving the Pfaffian invariant.
Numerically, we first construct the open-boundary Hamiltonian $H_{\rm OBC}$ for a fixed disorder realization. The boundary hopping matrix connecting the two ends of the strip is denoted by $T$. The periodically repeated Hamiltonian is then
\begin{equation}
H(q)=H_{\rm OBC}+e^{iq}T+e^{-iq}T^\dagger ,
\end{equation}
with the two particle-hole invariant twists
\begin{equation}
q=0,\pi .
\end{equation}
Thus,
\begin{align}
\begin{split}
H(0)=H_{\rm OBC}+T+T^\dagger ,
\\
H(\pi)=H_{\rm OBC}-T-T^\dagger .
\end{split}
\end{align}
These two cases correspond respectively to periodic and anti-periodic boundary conditions, or equivalently to zero and $\pi$ flux through the corresponding superconducting ring.
The Pfaffian periodic-disorder invariant used in this work is
\begin{equation}\label{eq.pdi}
\mathcal{Q}_{\rm PDI} = {\rm sgn} \left[ {\rm Pf}\,B(0)\, {\rm Pf}\,B(\pi)\right],
\end{equation}
where $B(0)=H(0)U_{\mathcal C}$, and $B(\pi)=H(\pi)U_{\mathcal C}$.
We use the convention
\begin{equation}
\mathcal{Q}_{\rm PDI}=-1
\end{equation}
for the topologically non-trivial phase and
\begin{equation}
\mathcal{Q}_{\rm PDI}=+1
\end{equation}
for the trivial phase. 
\textcolor{black}{Here, $\mathcal{Q}_{\rm PDI}$ denotes the Pfaffian-based periodic-disorder invariant of the periodically repeated disordered supercell, with $\mathcal{Q}_{\rm PDI}=-1$ and $\mathcal{Q}_{\rm PDI}=+1$ corresponding to nontrivial and trivial topological sectors, respectively. We emphasize that $\mathcal{Q}_{\rm PDI}$ is rigorously defined as a $\mathbb{Z}_2$ topological invariant of the periodically repeated system obtained by repeating a finite disordered graphene segment. As already mentioned, this type of construction restores translational symmetry at the supercell level and allows the topology of the corresponding bulk Bogoliubov-de Gennes Hamiltonian to be characterized through the Pfaffian formalism.
The physical systems studied in this work, however, are finite disordered graphene strips with open boundaries. Since the periodically repeated supercell preserves the same disorder realization and microscopic parameters as the finite strip, $\mathcal{Q}_{\rm PDI}$ probes the bulk topological character associated with the corresponding finite system. Through the bulk--boundary correspondence, a nontrivial Pfaffian sector indicates the tendency to support Majorana boundary modes. Nevertheless, finite-size hybridization, boundary effects, multichannel mixing, and partially separated Andreev bound states can modify the low-energy edge-state structure without changing $\mathcal{Q}_{\rm PDI}$. \textit{Therefore, when interpreted in the context of the finite graphene strips studied here, $\mathcal{Q}_{\rm PDI}$ is more appropriately regarded as a Pfaffian-based topological indicator, which we analyze together with Majorana polarization, low-energy spectra, and real-space wavefunction morphology.}
\textcolor{black}{In the numerical implementation, we explicitly verify the Hermiticity of $H(0)$ and $H(\pi)$, the particle-hole-symmetry residual, and the antisymmetry condition $B^{T}(\phi)=-B(\phi)$ prior to evaluating the Pfaffians. This verification is essential because the Pfaffian is defined only for antisymmetric matrices. The individual signs of ${\rm Pf},B(0)$ and ${\rm Pf},B(\pi)$ depend on the chosen basis convention, whereas their product yields the physically meaningful $\mathbb{Z}_2$ topological invariant $\mathcal{Q}_{\rm PDI}$.}
}}

\textcolor{black}{
Overall, we find that the nontrivial Pfaffian sector ($\mathcal{Q}_{\rm PDI}=-1$) generally correlates with regions exhibiting finite Majorana polarization and low-energy edge-localized states, particularly in the clean and weak-to-moderate disorder regimes. However, this correspondence is not strictly one-to-one in finite disordered systems. In several parameter regions, $\mathcal{Q}_{\rm PDI}=-1$ persists even when the corresponding low-energy states display partially separated Andreev-bound-state-like (psABS-like) wavefunction profiles due to finite-size overlap, multichannel mixing, or disorder-induced mode hybridization. Conversely, enhanced Majorana polarization can occur in regions containing a large density of disorder-induced low-energy states. These observations demonstrate that $\mathcal{Q}_{\rm PDI}$, Majorana polarization, low-energy spectra, and real-space wavefunction morphology provide complementary information and should be analyzed together when identifying candidate Majorana-supporting regimes in finite disordered graphene strips.}

%\textcolor{magenta}{Throughout the remainder of this manuscript, $\mathrm{Q}_{\rm PDI}$ should be understood as a Pfaffian-based topological indicator for the finite open-boundary graphene strips studied here, while $\mathcal{Q}_{\rm PDI}$ denotes the corresponding $\mathbb{Z}_2$ Pfaffian invariant of the periodically repeated disordered supercell for which the topological classification is rigorously defined.}
\textcolor{black}{
We conclude this section by emphasizing that the Pfaffian periodic-disorder invariant is most naturally defined for one-dimensional or quasi-one-dimensional class-D superconductors, where a preferred longitudinal direction allows a clear boundary-twist construction of the bulk topology. For nearly square geometries, the stronger two-dimensional confinement and increased multichannel mixing make this interpretation less direct. Accordingly, in square geometries, the Pfaffian analysis should be best regarded as an auxiliary qualitative indicator of topology included for comparison and completeness.
}

\section{Graphene Geometries}
Graphene in the bulk respects time-reversal, inversion, and threefold rotational symmetry ($C_3$), which together protect the local and global stability of its Dirac points~\cite{bernevig2013topological}. In contrast, finite-size realizations such as nanoribbons and flakes break some of these symmetries and display rich boundary physics, including edge-localized states and geometry-dependent spectral gaps. A key distinction arises from the termination: zigzag edges naturally host states pinned near zero energy, while armchair edges do not. When graphene is proximitized by an $s$-wave superconductor in addition to Rashba spin–orbit coupling and a Zeeman field, the nature of its edges plays a decisive role in shaping zero-energy Majorana bound states. We find that the stability of Majorana-supporting topological phases is not universal to all proximitized graphene structures, but depends critically on the edge termination and the orientation of the applied field.
This aspect will be explored in detail in the following sections. 

Finite-size graphene strips are formed by cutting a graphene sheet along specific crystallographic directions~\cite{son2006half,li2008chemically,jiao2009narrow,cai2010atomically}.  Fig.~\ref{f0} (a) illustrates a finite-size armchair graphene strip with $N = 5$ and $L = 15$ (A-strip), (b) zigzag strip with $N = 21$ and $L = 3$ (Z-strip), (c) square strip $N = 12$ and $L = 5$ (S-strip) where $N$ denotes the layer index and $L$ represents the unit cell. Note that the actual system size we consider in our study is much larger. \textcolor{black}{The labels A-, Z-, and S-type basically refer to different finite-size shape regimes of the same graphene honeycomb lattice. Specifically, A-type denotes the quasi-one-dimensional limit $N \ll L$, Z-type denotes $N \gg L$, and S-type denotes the nearly isotropic regime $N \sim L$, where the system dimensions along the two directions are comparable.}  

Specifically, we analyze three representative geometries: \textcolor{black}{a Z-strip with $N=170$ and $L=3$}, an A-strip with $N=3$ and $L=171$, and an S-strip with $N=35$ and $L=15$. In each case, the total number of lattice sites is $N \times (2L)$. Throughout this work, the strips are oriented such that the armchair edge lies along the $x$ direction.

\section{Majorana Zero Modes In Graphene Strips}
\begin{figure*}
    \centering
    \includegraphics[width=1.99\columnwidth]{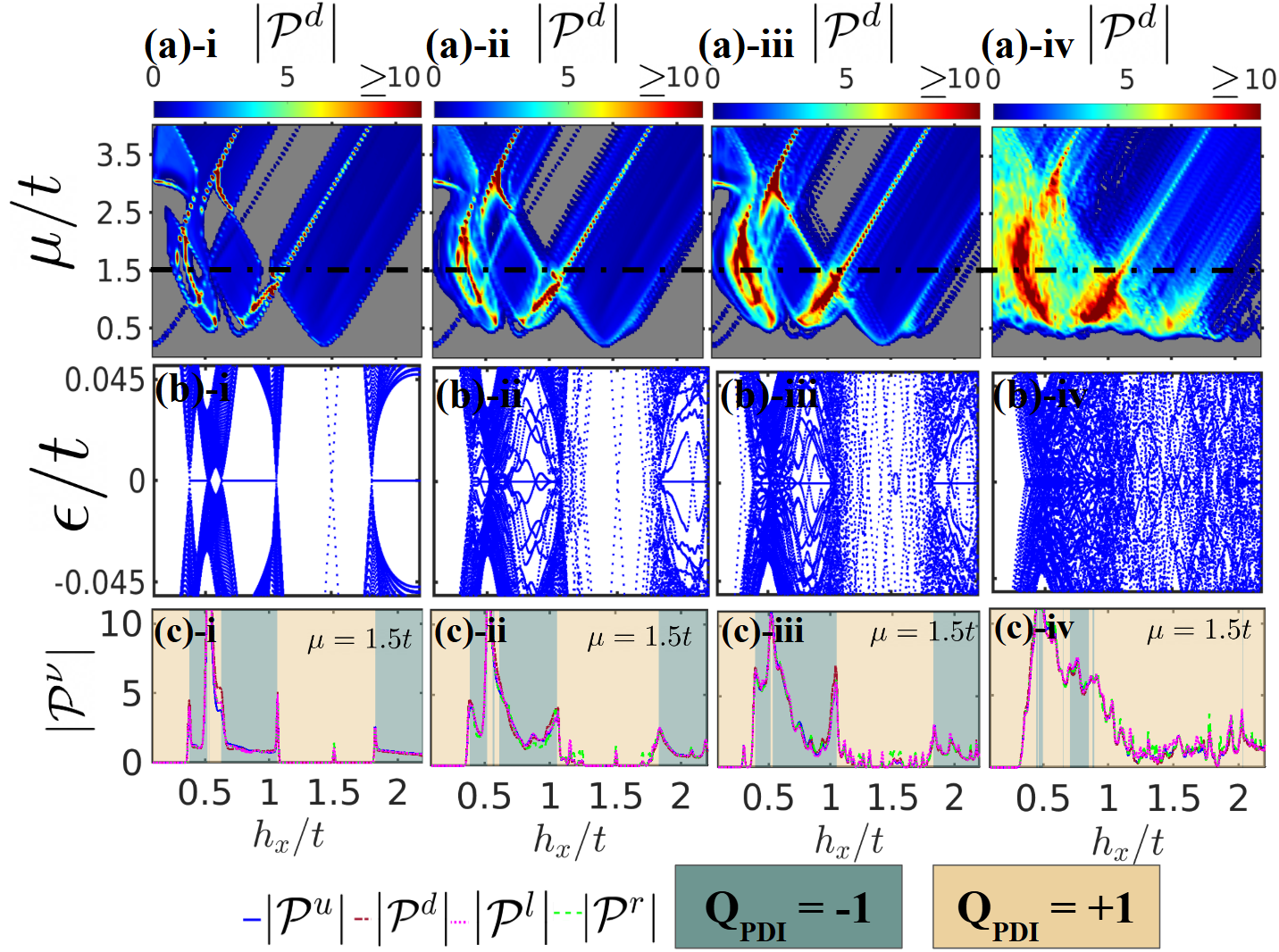}
    %\caption{

\caption{\textcolor{black}{Majorana polarization and low-energy spectrum in a finite-size \textbf{armchair} graphene strip under an in-plane magnetic field. Each column (i--iv) corresponds to increasing disorder strength: (i) clean system (no disorder), (ii) weak disorder ($\mathcal{Z}=20$), 20 impurities, $V_0=1.0$, (iii) moderate disorder ($\mathcal{Z}=50$), 50 impurities, $V_0=1.0$, and (iv) strong disorder ($\mathcal{Z}=150$), 150 impurities, $V_0=1.5$. Panels: (a) Heatmap of the absolute value of the representative Majorana polarization in the lower half of the strip, $|\mathcal{P}^{d}|$, as a function of Zeeman energy ($h_x/t$) and chemical potential ($\mu/t$). Since $|\mathcal{P}^{u}|$, $|\mathcal{P}^{d}|$, $|\mathcal{P}^{l}|$, and $|\mathcal{P}^{r}|$ display qualitatively similar behavior, only $|\mathcal{P}^{d}|$ is shown in the full parameter scan. The horizontal black dashed-dotted line indicates the representative chemical potential cut used for the low-energy spectrum and regional polarization plots shown in panels (b) and (c). Gray regions denote negligible polarization. (b) Low-energy spectrum (lowest 100 eigenvalues) as a function of $h_x/t$ at fixed chemical potential $\mu=1.5$. (c) Absolute value of the four regional Majorana polarizations $|\mathcal{P}^{\nu}|$ with $\nu=u,d,l,r$ (upper, lower, left, right halves), plotted as a function of $h_x/t$ at the same fixed chemical potential $\mu=1.5$ used in panel (b). Regions of enhanced polarization should be interpreted together with the spectral and real-space wavefunction diagnostics when identifying candidate Majorana zero modes. The background shading in panels (c) denotes the Pfaffian topological sector obtained from the periodic disorder \textcolor{black}{indicator} calculation, with $\mathrm{Q}_{\mathrm{PDI}}=-1$ (teal) and $\mathrm{Q}_{\mathrm{PDI}}=+1$ (beige) corresponding to nontrivial and trivial sectors, respectively.}}
    \label{fig:A_X}
\end{figure*}

Majorana zero modes useful for topological quantum computation are distinguished by the following features: (a) zero-energy protected by the superconducting spectral gap, (b) a non-zero Majorana polarization, and (c) localized wave functions at the edges or ends of the graphene strip. In the following subsections, we investigate topological superconductivity in graphene strips with armchair, zigzag, and square geometries. True Majorana modes are identified using the three criteria outlined above, after which we examine how disorder modifies their stability. To assess robustness, we consider three disorder regimes: (i) weak disorder with 20 impurities ($\mathcal{Z}=20$) and $V_0=1.0$, (ii) moderate disorder with 50 impurities ($\mathcal{Z}=50$) and $V_0=1.0$, and (iii) strong disorder with 150 impurities ($\mathcal{Z}=150$) and $V_0=1.5$. Representative spatial profiles of the disorder potential are shown in Fig.~\ref{fig:potential_profile}. In contrast to the simple one-dimensional nanowire, where the criterion $h>\sqrt{\mu^2+\Delta^2}$ sharply defines the topological region, the corresponding phase structure in finite graphene strips is more intricate, non-universal, and geometry-dependent. Furthermore, multiple overlapping polarization triangles are observed in the phase diagram as $h$ and $\mu$ are varied, which is a generic feature of multiband systems~\cite{sedlmayr2015visualizing,sedlmayr2016majorana,kaladzhyan2017majorana,kaladzhyan2017formation}. As the bandstructure varies upon changing parameters, one or more bands cuts across the chemical potential causing the appearance and disappearance of topologically protected Majorana bound states.  
\textcolor{black}{To systematically identify candidate Majorana zero modes in the different graphene geometries, we employ a hierarchical diagnostic procedure combining polarization, spectral, and real-space information. For each geometry, Figs.~\ref{fig:A_X}, \ref{fig:Z_Y}, and \ref{fig:S_Z}(a) provide two-dimensional scans of the representative Majorana polarization over the chemical-potential and Zeeman-field parameter space, which serve as an initial screening tool to locate regions where low-energy states acquire appreciable Majorana-like character. The representative chemical potentials selected for further analysis are indicated in these phase diagrams by horizontal black dashed-dotted lines. Guided by these maps, we examine the corresponding low-energy spectra and regional polarizations as a function of magnetic field in Figs.~\ref{fig:A_X}, \ref{fig:Z_Y}, and \ref{fig:S_Z}(b,c). This second step allows us to identify gap closings/reopenings, isolated near-zero-energy modes, and the persistence of finite polarization. Finally, for selected parameter points within these candidate regions, the real-space probability distributions shown in Figs.~\ref{fig:A_X_wf}, \ref{fig:Z_Y_wf}, and \ref{fig:S_Z_wf} are analyzed to distinguish opposite-edge localized states consistent with Majorana bound states from trivial bulk-localized, single-edge, or partially separated Andreev bound states.
%In these wavefunction figures, the panel labels begin from (d) in order to continue the sequential labeling scheme established by panels (a)--(c) of the preceding spectral/polarization figures. 
Throughout this work, no single observable is used in isolation; rather, the conclusions are based on the combined consistency of these complementary diagnostics \cite{karoliya2025majorana}.}

\textcolor{black}{Before discussing the individual geometries, we briefly comment on the dependence of the results on the direction of the applied magnetic field, which can be understood from the quasi-one-dimensional character of the strip geometries and the directional structure of the Rashba spin-orbit coupling. In the A- and Z-type strips, the large aspect ratio leads to dominant electronic propagation along the longer direction, while the transverse direction is strongly confined and gives rise to quantized subbands. The Rashba interaction produces an effective in-plane spin texture perpendicular to the dominant propagation direction. Consequently, a Zeeman field applied along the ribbon axis becomes transverse to the dominant Rashba spin texture and can more effectively open the helical gap required for a topological superconducting phase. In contrast, a field applied along the transverse direction is approximately parallel to the dominant Rashba spin texture of the low-energy modes, leading instead to stronger inter-subband mixing and suppressing the formation of clear gapped regions.}

\textcolor{black}{For example, for a low-energy mode propagating along an in-plane direction
\[
\hat{\mathbf e}_{\parallel}
=
(\cos\theta,\sin\theta,0),
\]
the Rashba spin axis is
\[
\hat{\mathbf e}_{R}
=
\hat{\mathbf z}\times \hat{\mathbf e}_{\parallel}
=
(-\sin\theta,\cos\theta,0).
\]
Projecting onto a transverse subband gives the effective one-dimensional normal-state Hamiltonian
\[
H^{(n)}(k_{\parallel})
=
\xi_n(k_{\parallel})\sigma_0
+
\alpha_n k_{\parallel}
\left(\hat{\mathbf e}_{R}\cdot\boldsymbol\sigma\right)
+
\mathbf h\cdot\boldsymbol\sigma .
\]
Only the component of \(\mathbf h\) perpendicular to \(\hat{\mathbf e}_{R}\) opens the helical gap at the Rashba crossing. The component parallel to \(\hat{\mathbf e}_{R}\) mainly shifts the Rashba-split branches and does not efficiently generate the spinless helical regime needed for the effective p-wave superconducting phase.}

\subsection{Armchair}
In this subsection, we examine the system the A-type (armchair) graphene strip. For our calculations, the strip dimensions are chosen as $N = 3$ and $L = 171$, with an in-plane magnetic field applied along the $x$-direction, {i.e., along the longer direction}. The results are presented in Fig.~\ref{fig:A_X} in three panels ((a), (b), and (c)), with a set of four figures each, labeled (i)-(iv), corresponding to different disorder strengths: (i) a clean system, (ii) a weakly disordered system with 20 impurities and $V_0 = 1.0$, (iii) a moderately disordered system with 50 impurities and $V_0 = 1.0$, and (iv) a strongly disordered system with 150 impurities and $V_0 = 1.5$. Each panel gives following information: (a) the phase diagram showing the Majorana polarization as a function of Zeeman field $h_x$ and chemical potential $\mu$; (b) the energy spectrum, where the lowest 100 eigenvalues are plotted as a function of $h_x$ for a fixed $\mu$; and (c) the Majorana polarization $|\mathcal{P}^{\nu}|$ (with $\nu =$ u, d, l, r), plotted as a function of $h_x$ for the same value of $\mu$. To compute $|\mathcal{P}^{\nu}|$, \textcolor{black}{we define the region $\mathcal{R}$}~[\ref{Eq:Pregion1}] as comprising the upper half (u), lower half (d), left half (l), and right half (r) of the strip. The polarization $|\mathcal{P}^{\nu}|$ is obtained by summing the weighted contributions from the lowest 100 eigenmodes, within each spatial region. In our results, we find that $|\mathcal{P}^u|$, $|\mathcal{P}^d|$, $|\mathcal{P}^l|$, and $|\mathcal{P}^r|$ exhibit nearly same values. Therefore, for clarity and simplicity, only $|\mathcal{P}^d|$ is shown in the phase plots for all cases.
\begin{figure*}
    \centering
    \includegraphics[width=1.99\columnwidth]{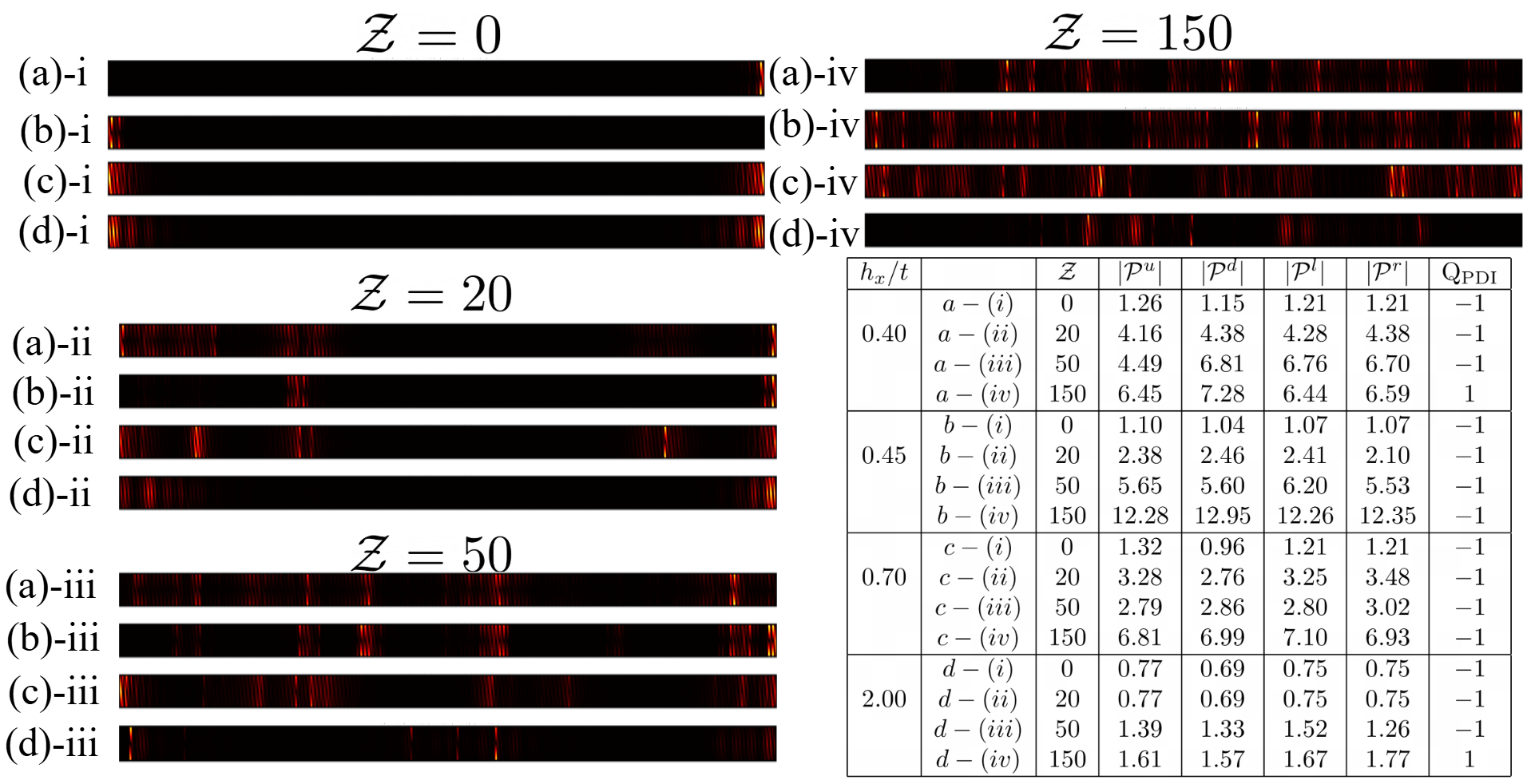}
    \caption{Real-space probability distribution of the low-energy Majorana modes on an \textbf{armchair} graphene nanoribbon under varying magnetic field and disorder strength. Panels (a)-(d) correspond to increasing values of Zeeman field $h_x/t = 0.40,\ 0.45,\ 0.70,$ and $2.0$, respectively, at fixed chemical potential $\mu/t = 1.5$. Each row (i-iv) shows the effect of increasing disorder strength: (i) Clean (no disorder), (ii) Weak disorder ($\mathcal{Z} = 20,\ V_0 = 1.0$), (iii) Moderate disorder ($\mathcal{Z} = 50,\ V_0 = 1.0$), and (iv) Strong disorder ($\mathcal{Z} = 150,\ V_0 = 1.5$). The adjacent table presents the absolute value of the Majorana polarization $|\mathcal{P}|$ for the lower ($\mathcal{P}^d$), upper ($\mathcal{P}^u$), left ($\mathcal{P}^l$), and right ($\mathcal{P}^r$) halves of the strip for each case. \textcolor{black}{The final column of the adjacent table lists the Pfaffian \textcolor{black}{indicator} sector corresponding to each parameter set.}}
    \label{fig:A_X_wf}
\end{figure*}

\textcolor{black}{Fig.~\ref{fig:A_X}(a) shows that regions with nonzero Majorana polarization appear as prominent band-like structures. The gray background corresponds to parameter regimes with negligible $\mathcal{P}^{d}$ and is used to enhance the visibility of finite-polarization regions. As the disorder strength increases from (a-i) to (a-iv), these structures broaden and gradually lose their sharp definition under strong disorder, indicating the degradation of clean candidate topological regions and the increasing presence of disorder-induced low-energy states.}
\textcolor{black}{The polarization maps in Fig.~\ref{fig:A_X}(a) should be interpreted as screening diagnostics rather than standalone topological phase diagrams. Regions of enhanced $|\mathcal{P}^{d}|$ indicate parameter windows where low-energy states acquire appreciable Majorana-like particle-hole character. With increasing disorder, the visibly reduced gray background and the apparent expansion of finite-polarization regions should not be interpreted as a growth of the topological phase. Rather, disorder generates additional near-zero-energy states and increases spectral mixing, so that more low-energy modes contribute finite particle-hole overlap to the polarization measure. While such states may display nonzero polarization, they need not correspond to genuine Majorana zero modes. \textit{For this reason, finite polarization is treated here as a necessary but not sufficient indicator, and candidate regions are further examined through the presence of a finite excitation gap and edge-localized real-space wavefunctions.}}
%{because from the figure it is clear that sharp peaks in the Majorana polarization (MP) plots coincide with the band closing and reopening events observed in the low-energy spectra, signaling topological phase transitions. Upon introducing strong disorder, the low-energy spectra no longer exhibit well-defined gap closing and reopening, and the MP plots become highly fluctuating. This behavior indicates the absence of a robust topological phase, demonstrating that strong disorder drives a breakdown of the topological phase in the system.} 
In panel (b), which presents the energy spectrum, we fix the chemical potential and plot the lowest 100 eigenvalues as a function of the Zeeman field $h_x$. For the clean system [Fig.~\ref{fig:A_X}(b-i)], clear band openings and closings are observed. Specifically, three distinct regions appear where zero-energy modes coexist with a finite spectral gap, which is a necessary, though not sufficient, condition for the presence of topologically protected Majorana modes. In the disordered cases, we observe that increasing disorder leads to the accumulation of low-energy states near zero energy. At high disorder [Fig.~\ref{fig:A_X}(b-iv)], the gap completely closes, indicating the absence of a topological phase. Up to moderate disorder, however, zero-energy states persist within partially open bands. To determine which of these regions correspond to genuine topological phases, we examine the Majorana polarization plots in the third panel, where $\mathcal{P}^\nu$'s are shown as a function of $h_x$ at $\mu = 1.5$ and wavefunction plots in the Fig.~\ref{fig:A_X_wf}. It is worth noting that sharp peaks in the Majorana polarization in Fig.~\ref{fig:A_X}(c) are typically associated with closing and reopening of the bulk gap, signaling topological phase transitions. \textcolor{black}{We note that the overall magnitude of the polarization may increase with disorder due to the growing density of near-zero-energy states, including trivial localized modes. This further highlights that finite polarization alone is not a sufficient indicator of topological Majorana modes.}    

We next turn to the spatial profiles of the wave functions, obtained by constructing a weighted superposition of the low-energy eigenstates, in the A-type (armchair) geometry. We observe a variety of localization behaviors, such as: (i) states localized at both ends of the strip, (ii) states localized only at a single end of the strip, (iii) states extended throughout the strip, (iv) states localized at intermediate positions on the strip away from the ends. %Notably, these features can occur even in cases with non-zero Majorana polarization and across different disorder regimes and geometries. 
In the wavefunction plots (Fig.~\ref{fig:A_X_wf}), the labels (a), (b), (c), etc., correspond to different values of the magnetic field $h_x/t$, while the sub-labels (i) - (iv) represent the increasing strength of the disorder: clean ($\mathcal{Z} = 0$), weak ($\mathcal{Z} = 20$), moderate ($\mathcal{Z} = 50$), and strong disorder ($\mathcal{Z} = 150$), respectively. Noting the spectrum in Fig.~\ref{fig:A_X} (b)-i, we identify three regions or \textit{phases} in $h_x/t$ where zero energy states appear with the spectral gap: (i) $h_x/t \approx 0.3$-$0.5$, (ii) $h_x/t \approx 0.55$-$1.0$, and (iii) $h_x/t \gtrsim 1.8$. We select representative values from each region to examine their wavefunction characteristics.

\textit{phase-(i):} In Fig.~\ref{fig:A_X_wf}, panels (a) and (b) correspond to $h_x/t = 0.40$ and $0.45$, respectively, both from the first phase. The wavefunctions in (a)-i and (b)-i are localized at only one edge of the strip, indicating that these zero-energy modes are topologically trivial. The role of disorder further supports this inference. In the low-energy spectra [Fig.~\ref{fig:A_X}(b)-ii, iii, iv], the first phase is no longer distinguishable as disorder increases; the corresponding zero-energy states merge into a continuum of low-energy modes, closing the bandgap necessary for topological protection. The associated wavefunctions [Fig.~\ref{fig:A_X_wf}(d)-ii-iv and (b)-ii-iv] lose their edge localization and become randomly distributed within the bulk of the strip, which suggests their trivial nature. 

\textit{phase-(ii) and (iii):} We next consider $h_x/t$ values from the second and third regions. In Fig.~\ref{fig:A_X_wf}(c)-i and (d)-i, corresponding to these phases, the wavefunctions are sharply localized at the zigzag (shorter) edges. To evaluate their robustness, we analyze both the low-energy spectra [Fig.~\ref{fig:A_X}(b)-ii–iv] and the wavefunctions [Fig.~\ref{fig:A_X_wf}(c)-ii-iii and (d)-ii-iii]. We find that these edge-localized modes enjoy gap-protection under weak and moderate disorder, although some additional bulk weight starts to appears with increasing disorder strength. Under strong disorder, however, the edge localization is lost, and the zero-energy modes vanish. These observations indicate that the second and third phases are topological in nature and can support robust Majorana zero modes, at least up to moderate levels of disorder. The table in the Fig.~\ref{fig:A_X_wf} has an absolute value for the left-, right-, upper-, and lower-half Majorana polarization values. 
\textcolor{black}{It is noteworthy that the Pfaffian \textcolor{black}{indicator} alone does not always guarantee well-separated Majorana-bound-state profiles in finite disordered systems. For example, the state shown in Fig.~\ref{fig:A_X_wf}(b-ii) lies within the nontrivial Pfaffian sector ($\mathrm{Q}_{\mathrm{PDI}}=-1$), yet its wavefunction exhibits psABS-like characteristics rather than sharply localized opposite-edge Majorana modes. This illustrates that the Pfaffian invariant \textcolor{black}{of the periodically repeated system} characterizes the global topology of the periodically repeated BdG Hamiltonian, whereas the real-space wavefunction morphology probes the degree of spatial separation, overlap, and finite-size hybridization of the corresponding low-energy boundary states. Consequently, global topology and local Majorana localization need not coincide perfectly in finite disordered graphene systems.
}

\subsection{Zigzag}
Here we examine the system with a Z-type (zigzag) graphene strip. To ensure the total number of lattice sites matches that of the A-type strip, \textcolor{black}{we set the strip dimensions to $N = 170$ and $L = 3$}. An in-plane magnetic field is applied along the $y$-direction, which is in the direction of the longer (zigzag) edge. 
\begin{figure*}
    \centering
    \includegraphics[width=1.99\columnwidth]{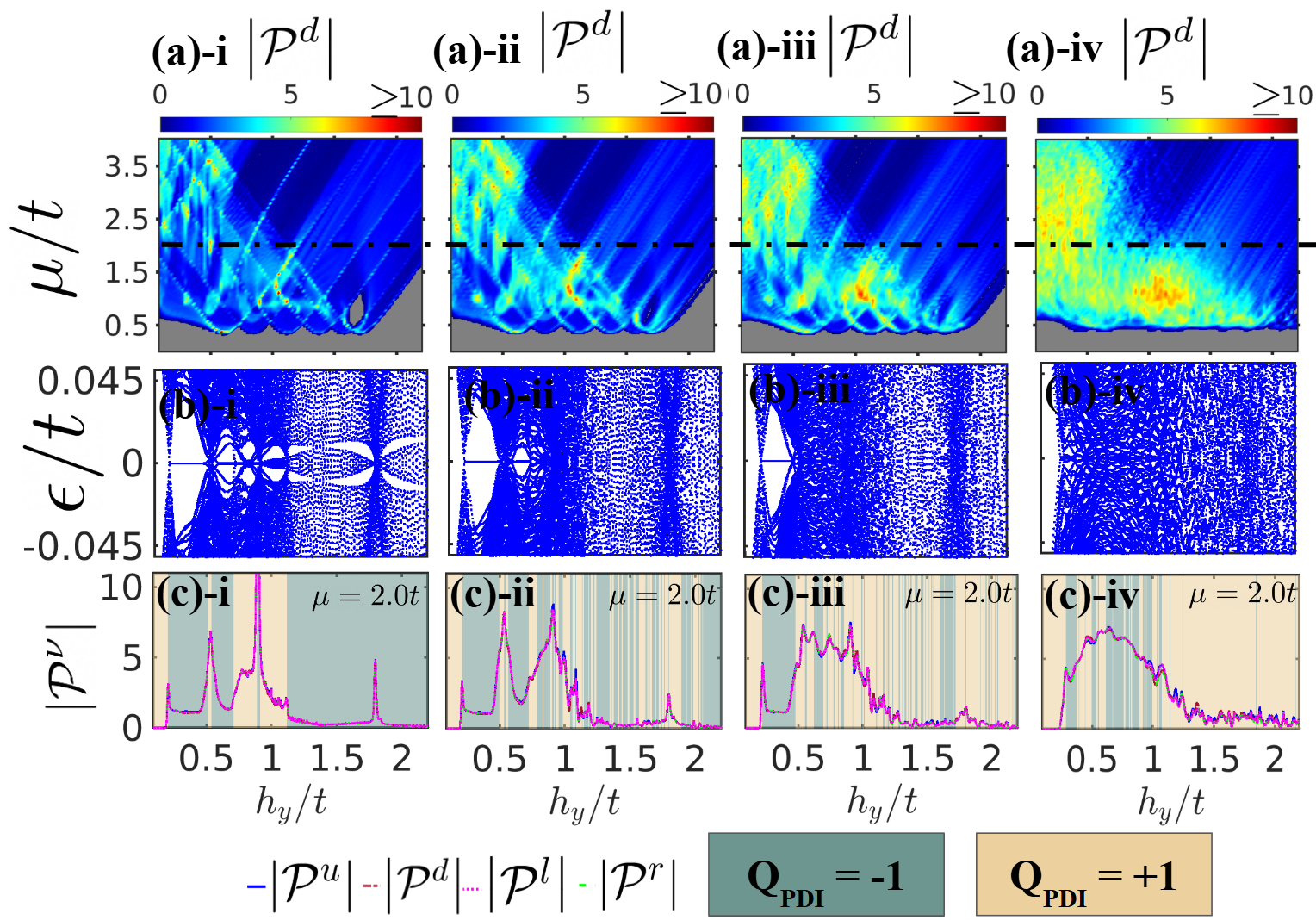}
    \caption{Majorana polarization in a finite-size \textbf{zigzag} graphene strip. Each column (i-iv) corresponds to a different disorder strength: 
(i) clean system (no disorder), 
(ii) weak disorder ($\mathcal{Z} = 20$), 20 impurities, $V_0 = 1.0$, 
(iii) moderate disorder ($\mathcal{Z} = 50$), 50 impurities, $V_0 = 1.0$, 
(iv) strong disorder ($\mathcal{Z} = 150$), 150 impurities, $V_0 = 1.5$.
Panels: 
(a) Heatmap of the absolute value of Majorana polarization in the lower half of the strip, $|\mathcal{P}^d|$, as a function of Zeeman energy ($h_y/t$) and chemical potential ($\mu/t$). \textcolor{black}{The horizontal black dashed-dotted line indicates the representative chemical potential cut used for the low-energy spectrum and regional polarization plots shown in panels (b) and (c).}
(b) Energy spectrum as a function of $h_y/t$ at fixed chemical potential $\mu = 2.0$. 
(c) Absolute value of Majorana polarization $|\mathcal{P}^\nu|$ with $\nu = u, d, l, r$ (upper, lower, left, right halves) as a function of $h_y/t$, also at $\mu = 2.0$. \textcolor{black}{The background shading in panels (c) denotes the Pfaffian topological sector obtained from the periodic disorder \textcolor{black}{indicator} calculation, with $\mathrm{Q}_{\mathrm{PDI}}=-1$ (teal) and $\mathrm{Q}_{\mathrm{PDI}}=+1$ (beige) corresponding to nontrivial and trivial sectors, respectively.}
}
    \label{fig:Z_Y}
\end{figure*}
The phase diagrams in Z-type strip shown in Fig.~\ref{fig:Z_Y}(a) exhibit a different structure compared to the A-type strip shown in Fig.~\ref{fig:A_X}(a).  For the Z-strip, we observe multiple sharp, V-shaped valley patterns formed by overlapping triangular regions, which can be contrasted to the A-type (armchair) case, where a few broad, overlapping triangular features dominate. These features are attributed to the different boundary conditions and sublattice terminations along the zigzag edge, which affect the localization and interference patterns of quasiparticle modes. As disorder increases, these sharp features become increasingly diffused and less well-defined. The second row [panels (b)-i to (b)-iv] shows the low-energy spectra at $\mu/t = 2.0$. \textcolor{black}{In contrast to the A-type system, where zero-energy modes persist over broader $h_x/t$ intervals, the Z-type strip features several narrow, discrete regions in $h_y/t$ where near-zero-energy states appear. This behavior reflects the stronger manifestation multiband nature of the finite zigzag strip, where different transverse subbands undergo successive gap closings and reopenings as the Zeeman field is varied.} \textcolor{black}{These parameter regions, which exhibit candidate topological signatures, become increasingly unstable under disorder:} as the disorder strength grows, the energy gaps close and non-topological low-energy states fill in, effectively destroying many of these narrow zero-energy windows. The third row [panels (c)-i to (c)-iv] presents the Majorana polarization as a function of $h_y/t$ for a fixed chemical potential $\mu/t = 2.0$, with polarization contributions plotted for four spatial regions of the strip. We note that in disordered systems as well, MP can be nonzero due to partially separated Andreev-bound states, which are not topologically protected.  
\begin{figure*}
    \centering
    \includegraphics[width=1.99\columnwidth]{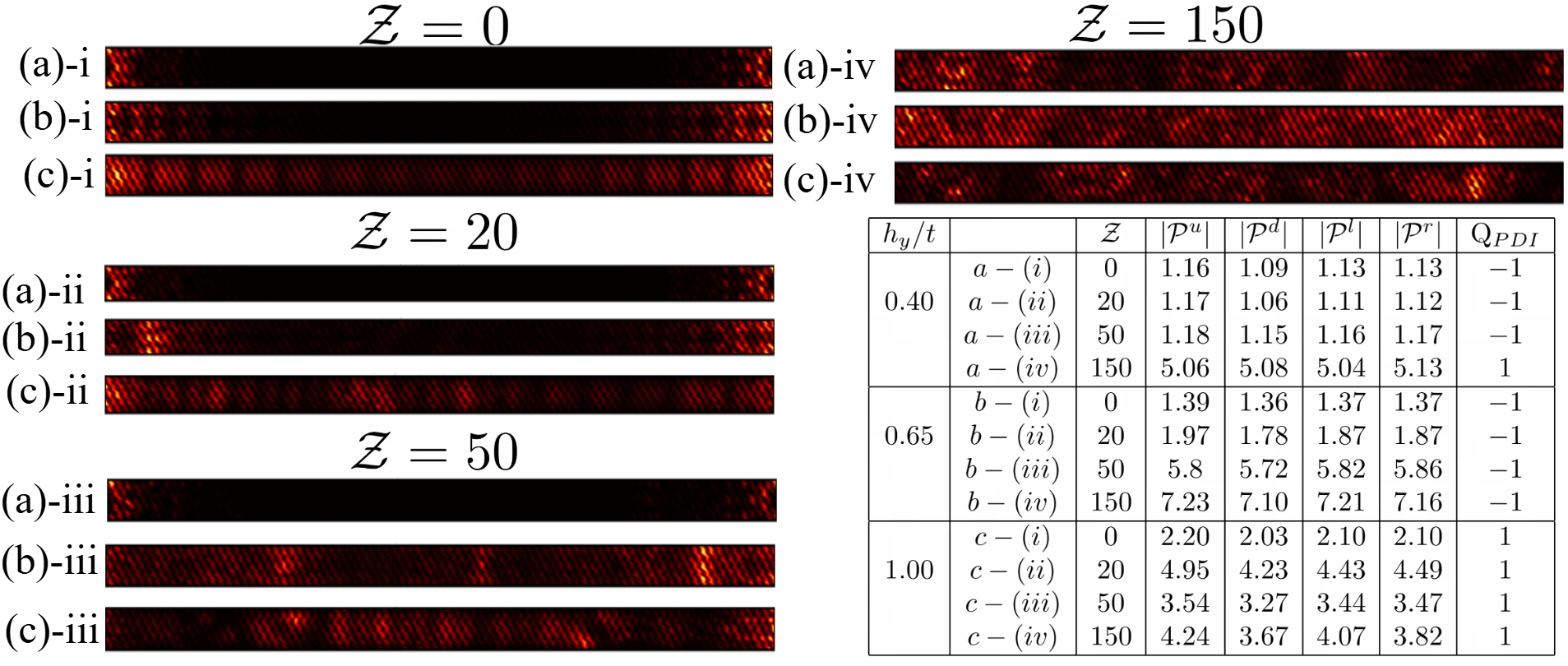}
    \caption{Real-space probability distribution of the low-energy Majorana modes on a \textbf{zigzag} graphene nanoribbon under varying magnetic field and disorder strength. Panels (a)-(c) correspond to increasing values of Zeeman field $h_y/t = 0.40,\ 0.65,$ and $1.00$, respectively, at fixed chemical potential $\mu/t = 2.0$. Each row (i-iv) shows the effect of increasing disorder strength: (i) Clean (no disorder), (ii) Weak disorder ($\mathcal{Z} = 20,\ V_0 = 1.0$), (iii) Moderate disorder ($\mathcal{Z} = 50,\ V_0 = 1.0$), and (iv) Strong disorder ($\mathcal{Z} = 150,\ V_0 = 1.5$). The adjacent table presents the absolute value of the Majorana polarization $|\mathcal{P}|$ for the lower ($\mathcal{P}^d$), upper ($\mathcal{P}^u$), left ($\mathcal{P}^l$), and right ($\mathcal{P}^r$) halves of the strip for each case. \textcolor{black}{The final column of the adjacent table lists the Pfaffian \textcolor{black}{indicator} sector corresponding to each parameter set.}}
    \label{fig:Z_Y_wf}
\end{figure*}

\textcolor{black}{To further investigate the existence and spatial character of Majorana modes in the Z-type strip, we analyze the low-energy wavefunctions corresponding to selected values of the Zeeman field $h_y/t$. From the low-energy spectra presented earlier [Fig.~\ref{fig:Z_Y}(b)], we identify two distinct parameter windows, approximately $h_y/t \in [0.22, 0.45]$ and $h_y/t \in [0.60, 0.70]$, where near-zero-energy states persist together with finite spectral gaps even in the presence of weak and moderate disorder, particularly for the first region. In a multiband quasi-one-dimensional system such as the present Z-type strip, different transverse subbands can undergo successive band inversions as the Zeeman field is varied, leading to reentrant trivial and topological regions separated by intermediate gap closings and reopenings}. To probe the nature of the zero-energy modes in these regions, we present wavefunction plots in Fig.~\ref{fig:Z_Y_wf} for $h_y/t = 0.40$, $0.65$, and $1.00$, each representative of a distinct phase. Panels (a)-i and (b)-i of Fig.~\ref{fig:Z_Y_wf} (clean system) show that for $h_y/t = 0.40$ and $0.65$, the zero-energy modes are localized at the shorter (armchair) edges of the strip. It is important to mention that edge localization in Z-type strips (shorter edges are armchair) is qualitatively different from that in A-type strips (shorter edges are zigzag). In the A-type system, the wavefunctions tend to exhibit sharp, confined peaks at the zigzag boundaries, whereas in the Z-type case, the intensity is more distributed along the edge, with a broader spatial profile. Panel (c)-i shows a similar edge-localized structure for $h_y/t = 1.00$, although with wavefunction intensity gradually decreasing from both ends toward the center. In the presence of weak disorder ($\mathcal{Z} = 20$), the edge-localized features for $h_y/t = 0.40$ and $0.65$ persist [panels (a)-ii, (b)-ii], indicating robustness of the corresponding candidate topological modes. \textcolor{black}{At moderate disorder ($\mathcal{Z} = 50$), however, we begin to observe the emergence of additional localized features in the bulk of the strip [panel (b)-iii], suggesting the coexistence of trivial modes. The gap closing observed near $h_y/t \simeq 0.5$ corresponds to the transition between the two distinct low-energy regions discussed above and should therefore be interpreted as an intermediate parity-changing transition rather than as the boundary between a single trivial and topological phase pair.} Under strong disorder ($\mathcal{Z} = 150$), the wavefunctions become fully delocalized or randomly localized, indicating that topological protection has been lost [panels (a)-iv, (b)-iv]. For $h_y/t = 1.00$, although the energy spectrum remains gapped, a high density of near-zero-energy states appears even in the clean limit. This is reflected in the wavefunction plots (c)-ii and (c)-iii, where the edge-localized features become less distinct and partially separated Andreev bound states (ps-ABS) emerge. These states are trivial low-energy states.
\textcolor{black}{
Again, a similar distinction between global topology and local wavefunction morphology is observed in the strongly disordered zigzag strip. In Fig.~\ref{fig:Z_Y_wf}(b-iv), the corresponding Pfaffian \textcolor{black}{indicator} remains nontrivial ($\mathrm{Q}_{\mathrm{PDI}}=-1$), while the low-energy wavefunction becomes substantially extended throughout the strip rather than remaining sharply localized near opposite edges. This behavior indicates that, although the periodically repeated BdG Hamiltonian retains a nontrivial topological sector, disorder-induced hybridization and multichannel mixing strongly degrade the spatial isolation of the low-energy boundary states in the finite system.
}

\subsection{Square}
After analyzing armchair (A-type) and zigzag (Z-type) graphene strips, we now turn to a system with nearly square geometry. We refer to it as “nearly square” due to anisotropies in the graphene lattice that prevent the geometry from being perfectly symmetric in both directions (see Fig.~\ref{f0} (c)). Along the vertical ($y$) direction, the distance between two adjacent A-sublattice sites lying on the same column is $\sqrt{3}a$, while along the horizontal ($x$) direction, the corresponding distance between A-sublattice sites on the same row is $3a$, where $a$ denotes the lattice constant. Thus, to approximate a square geometry while maintaining consistent spatial coverage and a comparable number of lattice sites to the armchair and zigzag systems, we choose $N = 35$ and $L = 15$. This configuration yields a geometry that is not strictly square but sufficiently close to allow a meaningful comparison of topological properties across different edge orientations, without introducing a significant bias due to system size.  

\begin{figure*}
    \centering
    \includegraphics[width=1.99\columnwidth]{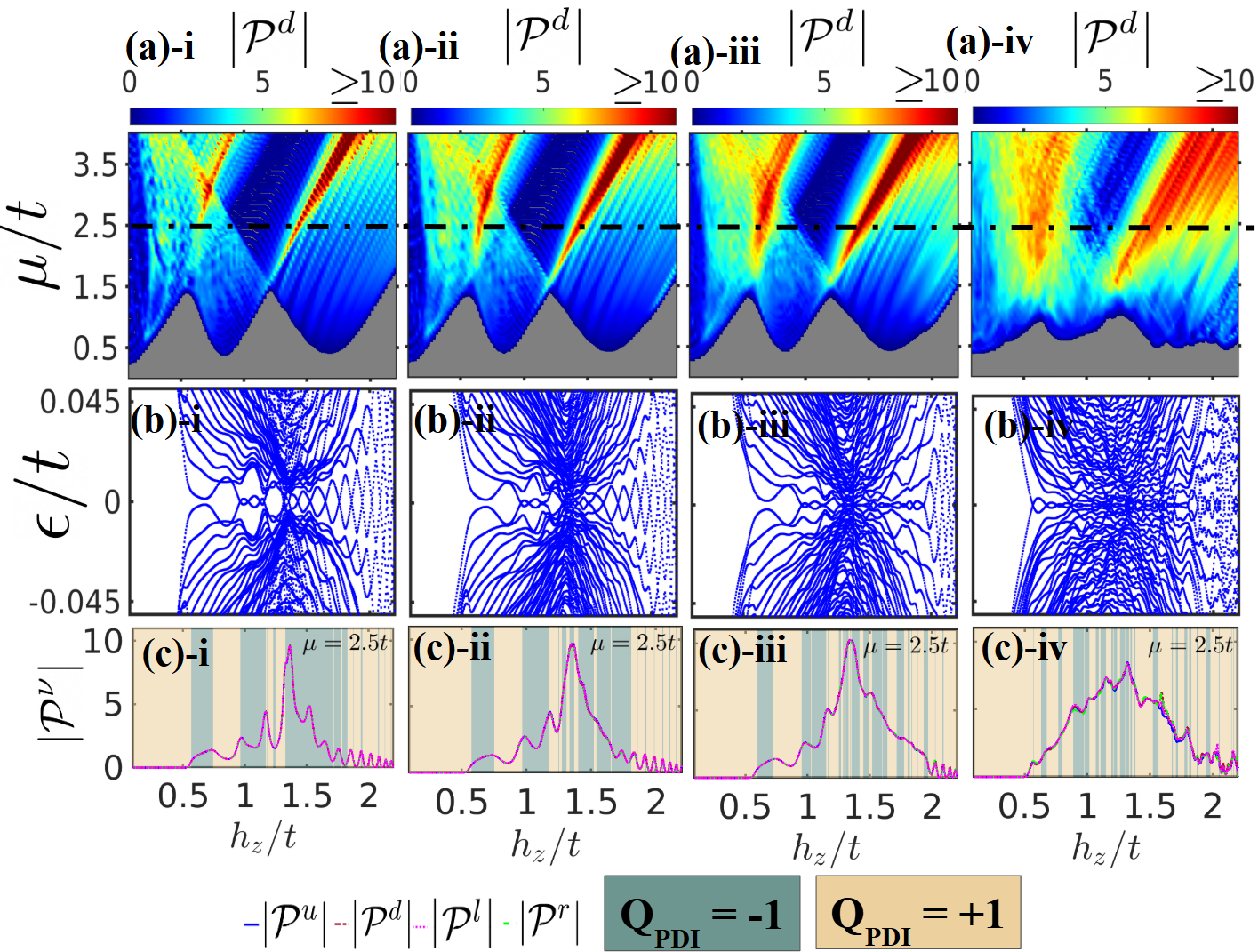}
    \caption{Majorana polarization in a finite-size \textbf{nearly square} shaped graphene geometry with \textbf{out-of-plane }magnetic field. Each column (i-iv) corresponds to a different disorder strength: 
(i) clean system (no disorder), 
(ii) weak disorder ($\mathcal{Z} = 20$), 20 impurities, $V_0 = 1.0$, 
(iii) moderate disorder ($\mathcal{Z} = 50$), 50 impurities, $V_0 = 1.0$, 
(iv) strong disorder ($\mathcal{Z} = 150$), 150 impurities, $V_0 = 1.5$.
Panels: 
(a) Heatmap of the absolute value of Majorana polarization in the lower half of the strip, $|\mathcal{P}^d|$, as a function of Zeeman energy ($h_z/t$) and chemical potential ($\mu/t$). \textcolor{black}{The horizontal black dashed-dotted line indicates the representative chemical potential cut used for the low-energy spectrum and regional polarization plots shown in panels (b) and (c).}
(b) Energy spectrum as a function of $h_z/t$ at fixed chemical potential $\mu = 2.5$. 
(c) Absolute value of Majorana polarization $|\mathcal{P}^\nu|$ with $\nu = u, d, l, r$ (upper, lower, left, right halves) as a function of $h_z/t$, also at $\mu = 2.5$.  \textcolor{black}{The background shading in panels (c) denotes the Pfaffian topological sector obtained from the periodic disorder \textcolor{black}{indicator} calculation, with $\mathrm{Q}_{\mathrm{PDI}}=-1$ (teal) and $\mathrm{Q}_{\mathrm{PDI}}=+1$ (beige) corresponding to nontrivial and trivial sectors, respectively.}
}
    \label{fig:S_Z}
\end{figure*}
If we apply an in-plane magnetic field, we find no clear phase separation or extended gapped regions.
\textcolor{black}
{
This behavior can be traced to the directional structure of the Rashba spin-orbit coupling. The Rashba term produces an in-plane spin-orbit field whose orientation depends on the direction of propagation. In a quasi-one-dimensional ribbon, the low-energy states are dominated by a single longitudinal direction. Thus, for the armchair strip with dominant modes propagating along the  $x$-direction, a $h_x\sigma_x$ Zeeman field is transverse to the dominant Rashba field, while for the zigzag strip with dominant
$y$-directed propagation, a $h_y\sigma_y$ field is transverse to the dominant Rashba field. In both cases, the in-plane field can open a gap.
In the nearly square flake, by contrast, low-energy states have both armchair-like and zigzag-like character. An in-plane Zeeman field cannot be transverse to the Rashba spin texture of all such modes simultaneously. It therefore gaps only part of the low-energy spectrum, while other modes remain weakly hybridized and continue to populate the midgap. As a result, the spectrum does not show a clean and well-separated gap for an in-plane Zeeman field.
\textcolor{black}{ A further important distinction arises from the confinement geometry itself. In the armchair and zigzag strips, the system is quasi-one-dimensional, with strong confinement primarily in a single transverse direction and relatively extended propagation along the long axis. This leads to a more well-defined subband structure and facilitates the isolation of low-energy channels capable of supporting cleaner topological gaps. By contrast, in the nearly square geometry, confinement occurs simultaneously in both spatial directions. Consequently, there is increased hybridization. This increased low-energy spectral density makes it more difficult to isolate sharply resolved zero-energy states and robust gapped regions, even when local Majorana-like signatures appear in the polarization and wavefunction profiles.}
An out-of-plane field $h_z\sigma_z$ is, however, transverse to the in-plane Rashba field for both propagation directions and hence gaps the spectrum more effectively.} 
We therefore primarily discuss the perpendicular-field configuration, i.e., with the Zeeman field applied along the $z$ direction. The first row [panels (a)-i to (a)-iv] of Fig.~\ref{fig:S_Z} displays the phase diagrams for the nearly square geometry under increasing disorder. The clean system [panel (a)-i] reveals a characteristic pattern of overlapping triangular regions in the $(\mu/t, h_z/t)$ parameter space, similar in structure to those seen in armchair and zigzag strips. {Figures~\ref{fig:S_Z}(a)-i,iv and \ref{fig:A_X}(a)-i,iv (or \ref{fig:Z_Y} (a)-i,iv) compare the $(\mu/t,h_z/t)$ phase diagrams for square and armchair (or zigzag) geometry, we can observe that for the square geometry, the largest distortion in the \textcolor{black}{promising topological phase} occur predominantly in the upper-right quadrant of the parameter space, corresponding to high chemical potential and high magnetic field. In contrast, for the armchair geometry, the \textcolor{black}{promising topological phase} is mainly concentrated in the low-field regime (upper-left and lower-left quadrants), with only faint signatures in the high-field region, as the disorder strength increases.}  

The second row [panels (b)-i to (b)-iv] presents the low-energy spectra, as a function of $h_z/t$ at fixed $\mu = 2.5$. In the clean limit [panel (b)-i], we observe gap closings and reopenings after $h_z/t \sim 0.7$, \textcolor{black}{{pointing towards promising topological phase transitions}}. However, unlike the armchair and zigzag cases, where extended regions of zero-energy states were observed, the square geometry exhibits more discrete, oscillatory zero-energy modes.
As disorder increases, the low-energy states begin to accumulate near zero energy, but lack a clear separation from the bulk spectrum. Under strong disorder, [panel (b)-iv], clear gap openings and reopenings are absent.
\begin{figure*}
    \centering
    \includegraphics[width=1.99\columnwidth]{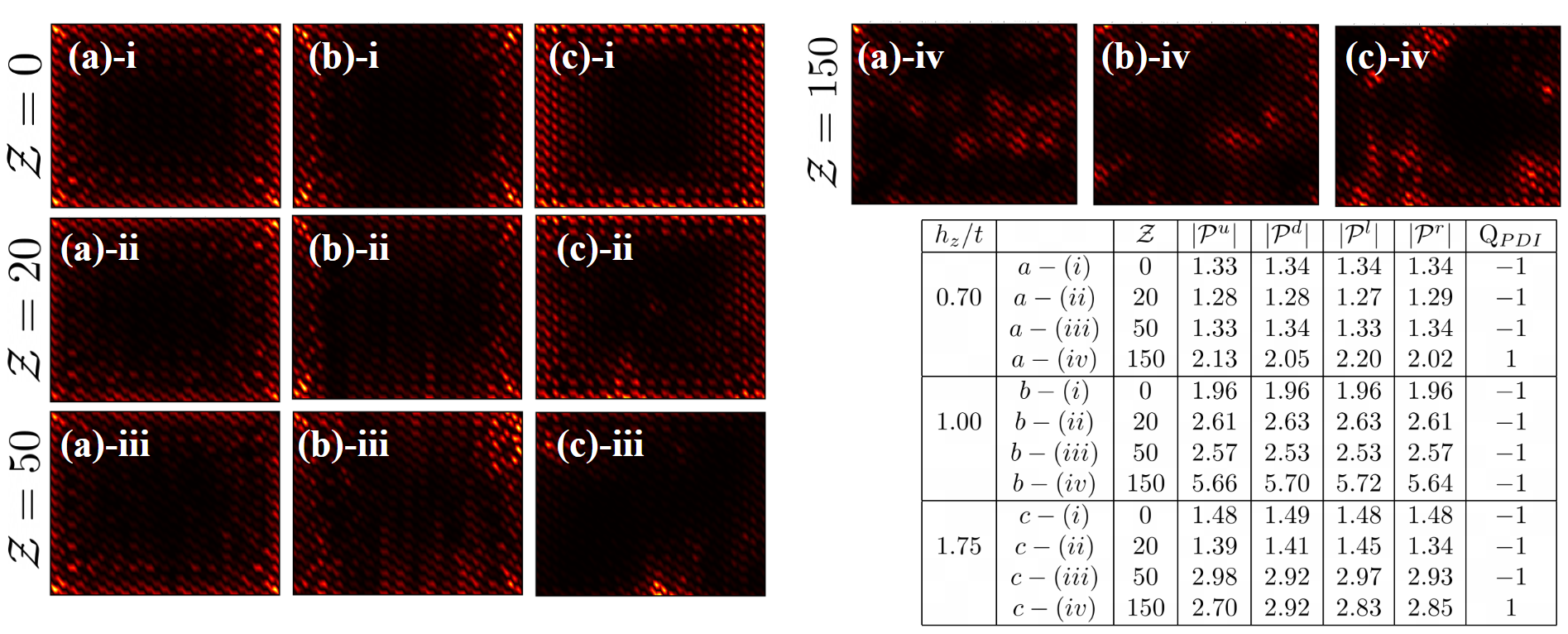}
    \caption{Real-space probability distribution of the low-energy Majorana modes on a nearly \textbf{square} graphene geometry under varying \textbf{out-of-plane} magnetic field and disorder strength. Panels (a)-(c) correspond to increasing values of Zeeman field $h_z/t = 0.70,\ 1.00,$ and $1.75$, respectively, at fixed chemical potential $\mu/t = 2.5$. Each row (i-iv) shows the effect of increasing disorder strength: (i) Clean (no disorder), (ii) Weak disorder ($\mathcal{Z} = 20,\ V_0 = 1.0$), (iii) Moderate disorder ($\mathcal{Z} = 50,\ V_0 = 1.0$), and (iv) Strong disorder ($\mathcal{Z} = 150,\ V_0 = 1.5$). The adjacent table presents the absolute value of the Majorana polarization $|\mathcal{P}|$ for the lower ($\mathcal{P}^d$), upper ($\mathcal{P}^u$), left ($\mathcal{P}^l$), and right ($\mathcal{P}^r$) halves of the strip for each case. \textcolor{black}{The final column of the adjacent table lists the Pfaffian \textcolor{black}{indicator} sector corresponding to each parameter set.}}
    \label{fig:S_Z_wf}
\end{figure*}
To further probe the nature of the low-energy states observed in Fig.~\ref{fig:S_Z}, we examine the spatial distribution of their corresponding wavefunctions in Fig.~\ref{fig:S_Z_wf}. These plots represent selected eigenstates at zero energy for three representative values of the Zeeman field: $h_z/t = 0.70$, $1.00$, and $1.75$, chosen from regions that exhibit moderate to high polarization and indications of \textcolor{black}{possible topological signatures}. Each row corresponds to increasing disorder strength: clean $\mathcal{Z} = 0$, $\mathcal{Z} = 20$, $\mathcal{Z} = 50$, and $\mathcal{Z} = 150$, respectively. In the clean system [panels (a)-i to (c)-i], the wavefunctions for $h_z/t = 0.70$ and $1.00$ show localization along the system boundaries, particularly at the corners and edges. This boundary-localized intensity, suggests the emergence of Majorana bound states (MBS) confined near the system's periphery. At $h_z/t = 1.75$ [panel (c)-i], although some edge intensity persists, the mode appears more spread and less sharply confined, indicating that this regime may correspond to a partially separated Andreev bound state (ps-ABS) rather than a topologically protected MBS. As disorder is introduced [panels (a–c)-ii and (a–c)-iii], the wavefunctions maintain edge-localized features at $h_z/t = 1.00$, \textcolor{black}{corresponding to a parameter regime exhibiting stronger Majorana-like characteristics. A detailed comparison between $h_z/t = 0.7$ and $h_z/t = 1.0$ shows that, while both exhibit near-zero-energy features, the latter displays reduced energy splitting, a non-zero polarization, and a more pronounced localization along opposite edges. This indicates a comparatively stronger Majorana-like character at $h_z/t = 1.0$, although the distinction remains quantitative rather than qualitative.} However, for $h_z/t = 0.70$ and $1.75$, the localization becomes increasingly fragmented, and the amplitude shifts toward interior regions of the strip, indicating that disorder weakens or destroys the topological character of these states more rapidly outside the $h_z/t \approx 1.00$ regime. In the strongly disordered case [panels (a–c)-iv], all wavefunctions lose clear spatial structure and exhibit random localization throughout the bulk. No clear edge confinement is visible, consistent with the mixing of low-energy states seen in the spectrum and the breakdown of the gapped regions in the phase diagrams. This confirms that the topological character of the system is lost under strong disorder, and the surviving zero-energy modes are trivial in nature.

\textcolor{black}{
We note that the Pfaffian-based periodic disorder invariant is fundamentally motivated by one-dimensional and quasi-one-dimensional class-D superconducting systems, where a well-defined ordering direction and boundary-twist construction naturally characterize the bulk topology. In the nearly square geometry considered here, the system exhibits stronger two-dimensional confinement and enhanced multichannel mixing, so the interpretation of the \textcolor{black}{indicator} becomes less direct than in the ribbon-like A- and Z-type strips. The Pfaffian analysis for the S-type geometry should therefore be regarded primarily as a complementary and qualitative topological diagnostic included for completeness and comparative purposes.
}

\subsection{Additional Cases} %% IN THIS SECTION, SHOULD WE MENTION QUADRANTWISE POLARIZATION CALCULATION? 
For completeness, we have also studied the A-type strip under a magnetic field applied along the $y$- and the Z-type strip under the field along the $x$-direction (i.e., field along the shorter directions). We also study both of them under a magnetic field in the $z$-direction as well. 
For the A-type geometry, we find that an in-plane field along the $y$-direction fails to produce any significant topological features: the phase diagrams remain distorted, and the low-energy spectra are dominated by states merged with the bulk, with no clear gapped regions. A similar lack of non-trivial behavior is observed in the Z-type strip when the magnetic field is applied in the $x$-direction. In both cases, the phase plots show no well-defined \textcolor{black}{structure to be a candidate for a topological phase}, and no isolated zero-energy modes are found. In contrast, the application of an out-of-plane ($z$-direction) magnetic field yields interesting topological behavior in both geometries. Since the essential qualitative features have already been discussed in earlier sections, the corresponding figures and discussions are relegated to the Appendix. 
\section{Conclusions and Outlook}
Although graphene-based platforms for realizing Majorana modes have been examined in some earlier works, a unified understanding that connects geometry, magnetic orientation, disorder, and Andreev-state correlations has been missing.
We revisited Majorana zero modes in proximitized graphene strips and examined under what conditions the observed zero-energy states correspond to genuinely nonlocal Majoranas rather than partially separated Andreev bound states (quasi-Majoranas). 
Building on earlier works that introduced Majorana polarization as a spatial diagnostic \cite{sticlet2012spin,sedlmayr2015visualizing}, we developed a systematic framework specific to graphene. Using a minimal tight-binding model that includes proximity-induced superconductivity, Zeeman coupling with tunable orientation, and site-dependent potential fluctuations (disorder), we correlated spectral evolution, MP distribution, and real-space wave-function morphology to identify regimes supporting robust, non-overlapping edge modes.
Specifically, we studied finite strips with armchair (A-type), zigzag (Z-type), and square (S) geometry, allowing us to track how different terminations modify the low-energy spectrum and spatial character of the bound states. By systematically analyzing all three geometries under magnetic fields oriented along the $x$-, $y$-, and $z$-axes, we find that both edge termination and field orientation critically determine the presence and robustness of Majorana modes. Among these, the A-type strips, with shorter zigzag edges, emerge as the most favorable hosts, exhibiting well-defined, gapped zero-energy phases with sharply localized edge states and finite Majorana polarization, even under moderate to strong disorder. In these systems, both in-plane ($x$) and out-of-plane ($z$) magnetic fields yield robust topological features. For Z-type strips, with shorter armchair edges, topological phases also emerge for both in-plane ($y$) and out-of-plane ($z$) magnetic fields. Here, zero-energy modes appear within comparable parameter ranges but exhibit slightly broader spatial profiles near the edges than in A-type strips, where localization is sharper. {In Fig.~\ref{f0}(a), which corresponds to an A-type strip with zigzag-terminated short edges, the two ends are sublattice polarized: the red and black dots denote the A and B sublattices, respectively, so the left boundary contains only A-sublattice sites, whereas the right boundary contains only B-sublattice sites. Such one-sublattice termination is characteristic of zigzag edges and gives rise to sublattice-polarized zero-energy states, in which the wavefunction has finite amplitude only on the terminating sublattice while the opposite component is forced to vanish by the tight-binding boundary conditions~\cite{nakada1996edge}. In contrast, armchair edges include both A and B sublattices at the boundary; the corresponding boundary condition mixes the two components rather than suppressing either of them. As a result, armchair edges do not support sublattice-polarized zero modes, and their low-energy states extend across the nanoribbon. These distinctions underlie the much sharper localization of Majorana bound states at zigzag edges compared to armchair edges in proximitized graphene nanoribbons.} In the S-type geometry, topological phases occur only when the magnetic field is applied along the $z$-direction. The corresponding wavefunctions display corner modes and edge localization along all four sides, with moderate spatial spread. These corner and edge states remain robust against weak to moderate disorder.
\textcolor{black}{To further strengthen the topological characterization, we additionally incorporated a Pfaffian-based periodic disorder \textcolor{black}{indicator} (PDI) analysis for the finite disordered graphene strips. The nontrivial Pfaffian sector generally correlates with regions exhibiting finite Majorana polarization, low-energy spectral features, and edge-localized states, particularly in the clean and weak-to-moderate disorder regimes. At the same time, our results demonstrate that global topology and local Majorana diagnostics are not strictly equivalent in finite disordered systems: parameter regions with nontrivial Pfaffian \textcolor{black}{indicator} may still exhibit partially separated Andreev-bound-state-like wavefunctions due to finite-size hybridization and multichannel mixing. These findings therefore, highlight the importance of combining global topological invariants with local spectral and wavefunction diagnostics when identifying robust Majorana-supporting regimes in proximitized graphene nanostructures.}
Looking forward, several open directions emerge. Investigating how the ground-state splitting scales with system length would further clarify the nonlocal nature of the modes. Incorporating self-consistent electrostatics and spatially varying pairing potentials could improve quantitative accuracy, while multiterminal configurations may reveal nonlocal conductance signatures unique to Majorana pairs. Our results elevate graphene from a conceptual setting to a controllable material platform where the interplay of geometry, magnetic orientation, and potential landscape can be exploited to engineer, stabilize, and systematically characterize Majorana-bound-state-supporting regimes.

%{In addition, recent progress on the periodic disorder invariant (PDI), which constructs a bulk topological invariant by periodically repeating a finite disordered segment \cite{eissele2025topological}, offers a promising future route for diagnosing disorder-robust Majorana zero modes and distinguishing them from trivial zero-energy states in proximitized graphene nanoribbons.}
\textit{Acknowledgment:} G.S. was funded by ANRF-SERB Core Research Grant CRG/2023/005628. S.K. was funded by IIT Mandi HTRA fellowship. S.T. acknowledges ARO Grant W911NF2210247. We thank Ekta for independently verifying our results using a KWANT-based numerical implementation.\\

\appendix
\begin{widetext}

\normalsize
\section{Armchair with out-of-plane Zeeman field}
\begin{figure}
 \centering
 \includegraphics[width=0.7\linewidth]{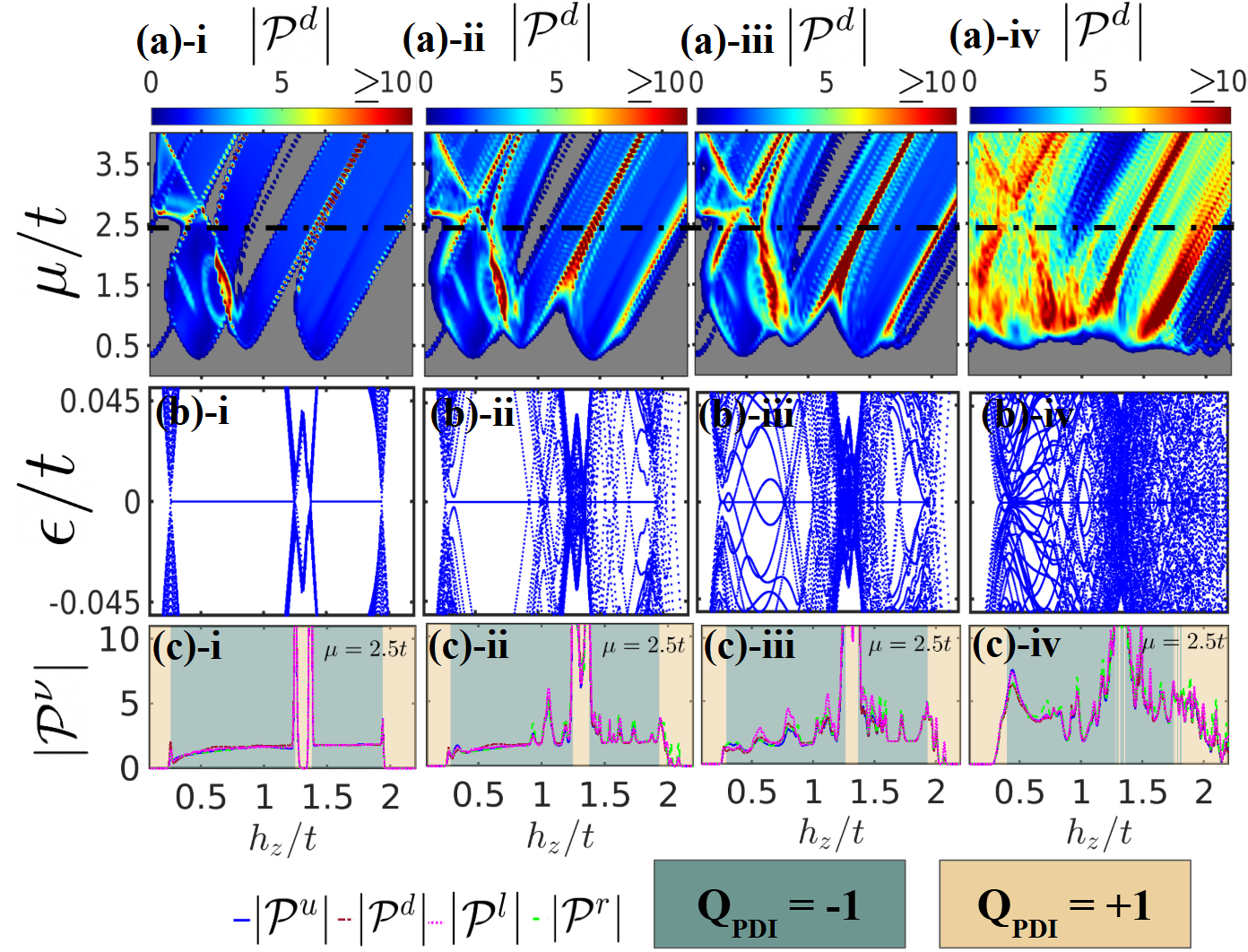}
 \caption{
Majorana polarization in a finite-size \textbf{armchair} graphene strip for \textbf{out-of-plane magnetic field}. Each column (i-iv) corresponds to a different disorder strength: 
(i) clean system (no disorder), 
(ii) weak disorder ($\mathcal{Z} = 20$), 20 impurities, $V_0 = 1.0$, 
(iii) moderate disorder ($\mathcal{Z} = 50$), 50 impurities, $V_0 = 1.0$, 
(iv) strong disorder ($\mathcal{Z} = 150$), 150 impurities, $V_0 = 1.5$.
Panels: 
(a) Heatmap of the absolute value of Majorana polarization in the lower half of the strip, $|\mathcal{P}^d|$, as a function of Zeeman energy ($h_z/t$) and chemical potential ($\mu/t$). \textcolor{black}{The horizontal black dashed-dotted line indicates the representative chemical potential cut used for the low-energy spectrum and regional polarization plots shown in panels (b) and (c).}
(b) Energy spectrum as a function of $h_z/t$ at fixed chemical potential $\mu = 2.5$. 
(c) Absolute value of Majorana polarization $|\mathcal{P}^\nu|$ with $\nu = u, d, l, r$ (upper, lower, left, right halves) as a function of $h_z/t$, also at $\mu = 2.5$. \textcolor{black}{The background shading in panels (c) denotes the Pfaffian topological sector obtained from the periodic disorder \textcolor{black}{indicator} calculation, with $\mathrm{Q}_{\mathrm{PDI}}=-1$ (teal) and $\mathrm{Q}_{\mathrm{PDI}}=+1$ (beige) corresponding to nontrivial and trivial sectors, respectively.}
}
 \label{fig:A_Z_1}
\end{figure}

\begin{figure}
 \centering
 \includegraphics[width=0.9\linewidth]{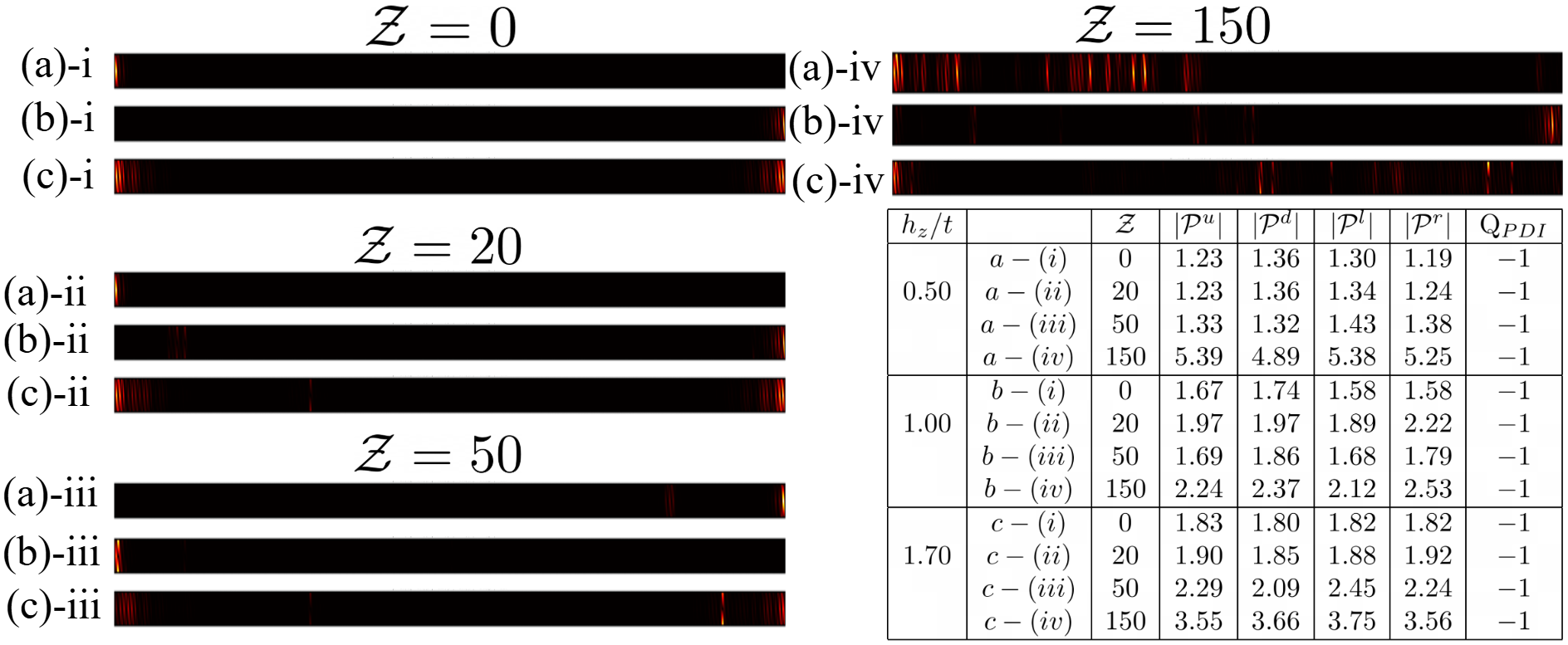}
 \caption{Real-space probability distribution of the low-energy Majorana modes on an \textbf{armchair} graphene nanoribbon under varying \textbf{out-of-plane magnetic field} and disorder strength. Panels (a)-(c) correspond to increasing values of Zeeman field $h_z/t = 0.50,\ 1.00,\ $ and $1.70$, respectively, at fixed chemical potential $\mu/t = 2.5$. Each row (i-iv) shows the effect of increasing disorder strength: (i) Clean (no disorder), (ii) Weak disorder ($\mathcal{Z} = 20,\ V_0 = 1.0$), (iii) Moderate disorder ($\mathcal{Z} = 50,\ V_0 = 1.0$), and (iv) Strong disorder ($\mathcal{Z} = 150,\ V_0 = 1.5$). The adjacent table presents the absolute value of the Majorana polarization $|\mathcal{P}|$ for the lower ($\mathcal{P}^d$), upper ($\mathcal{P}^u$), left ($\mathcal{P}^l$), and right ($\mathcal{P}^r$) halves of the strip for each case. \textcolor{black}{The final column of the adjacent table lists the Pfaffian \textcolor{black}{indicator} sector corresponding to each parameter set.}}
 \label{fig:A_Z_wf}
\end{figure}

We analyze the spectral and polarization plots for the armchair strip in the presence of disorder with an out-of-plane magnetic field, as shown in Fig.~\ref{fig:A_Z_1}. The low-energy spectra in panels (b)-i–iv reveal two distinct topological regions, where zero-energy modes persist when the magnetic field lies in the ranges $h_z/t \approx 0.4$–$1.2$ and $1.5$–$1.9$, with the chemical potential fixed at $\mu/t = 2.5$. Importantly, the bulk gap remains open up to moderate disorder, indicating robustness of the topological phase. Complementary evidence for topological regions comes from the polarization plots in panels (c)-i–iv, which display two sharp peaks characteristic of topological phase transitions. Within the above field ranges, the polarization remains stable in the clean system and exhibits only small oscillations as the disorder strength increases. Further insight is obtained from the wavefunction profiles in Fig.~\ref{fig:A_Z_wf}. Robust Majorana zero modes are clearly observed in panels (c)-i–iii, corresponding to clean to moderately disordered systems at $h_z/t = 1.70$. These Majorana modes are sharply localized at the strip ends. In contrast, panel (a)-iv reveals partially separated Andreev bound states, while panels (b)-i–iii display states localized at a single edge with finite polarization, signifying trivial modes. The phase diagram for the A-type geometry primarily exhibits a triangular pattern, which gradually distorts with increasing disorder at higher $\mu/t$ and $h_z/t$ values. The low-energy spectrum reveals multiple gapped regions supporting zero-energy modes, whose corresponding wavefunctions remain sharply localized at the zigzag (shorter) edges. This localization persists even under moderate to strong disorder, signaling the emergence of robust, topologically protected Majorana bound states (MBS). Occasionally, however, we find wavefunctions localized at a single edge, consistent with partially separated Andreev bound states (ps-ABS), which are topologically trivial.

\begin{figure}
 \centering
 \includegraphics[width=0.7\linewidth]{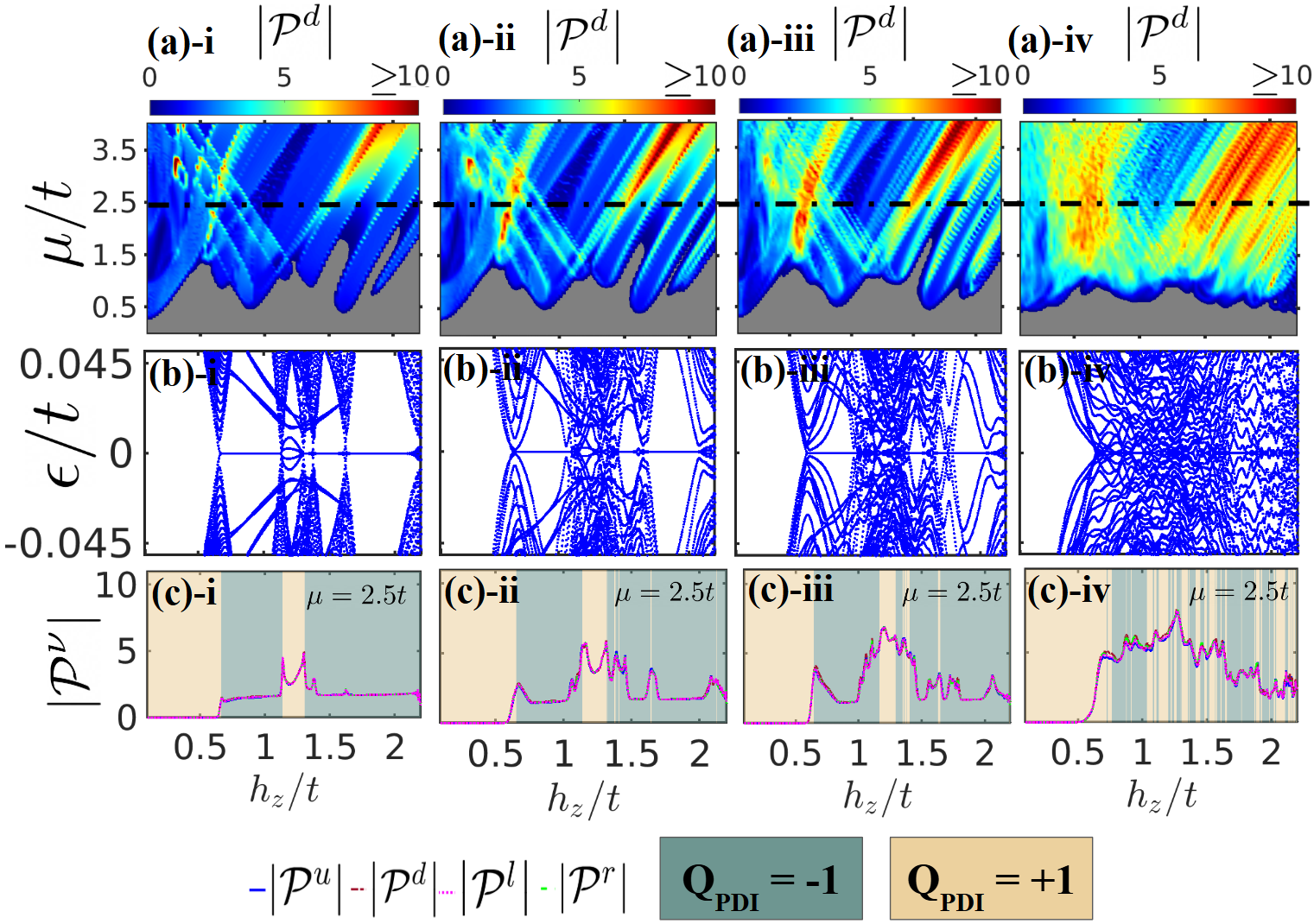}
 \caption{Majorana polarization in a finite-size \textbf{zigzag} graphene strip for \textbf{out-of-plane magnetic field}. Each column (i-iv) corresponds to a different disorder strength: 
(i) clean system (no disorder), 
(ii) weak disorder ($\mathcal{Z} = 20$), 20 impurities, $V_0 = 1.0$, 
(iii) moderate disorder ($\mathcal{Z} = 50$), 50 impurities, $V_0 = 1.0$, 
(iv) strong disorder ($\mathcal{Z} = 150$), 150 impurities, $V_0 = 1.5$.
Panels: 
(a) Heatmap of the absolute value of Majorana polarization in the lower half of the strip, $|\mathcal{P}^d|$, as a function of Zeeman energy ($h_z/t$) and chemical potential ($\mu/t$). \textcolor{black}{The horizontal black dashed-dotted line indicates the representative chemical potential cut used for the low-energy spectrum and regional polarization plots shown in panels (b) and (c).}
(b) Energy spectrum as a function of $h_z/t$ at fixed chemical potential $\mu = 2.5$. 
(c) Absolute value of Majorana polarization $|\mathcal{P}^\nu|$ with $\nu = u, d, l, r$ (upper, lower, left, right halves) as a function of $h_z/t$, also at $\mu = 2.5$. \textcolor{black}{The background shading in panels (c) denotes the Pfaffian topological sector obtained from the periodic disorder \textcolor{black}{indicator} calculation, with $\mathrm{Q}_{\mathrm{PDI}}=-1$ (teal) and $\mathrm{Q}_{\mathrm{PDI}}=+1$ (beige) corresponding to nontrivial and trivial sectors, respectively.}}
 \label{fig:Z_Z_1}
\end{figure}

\begin{figure}
 \centering
 \includegraphics[width=0.9\linewidth]{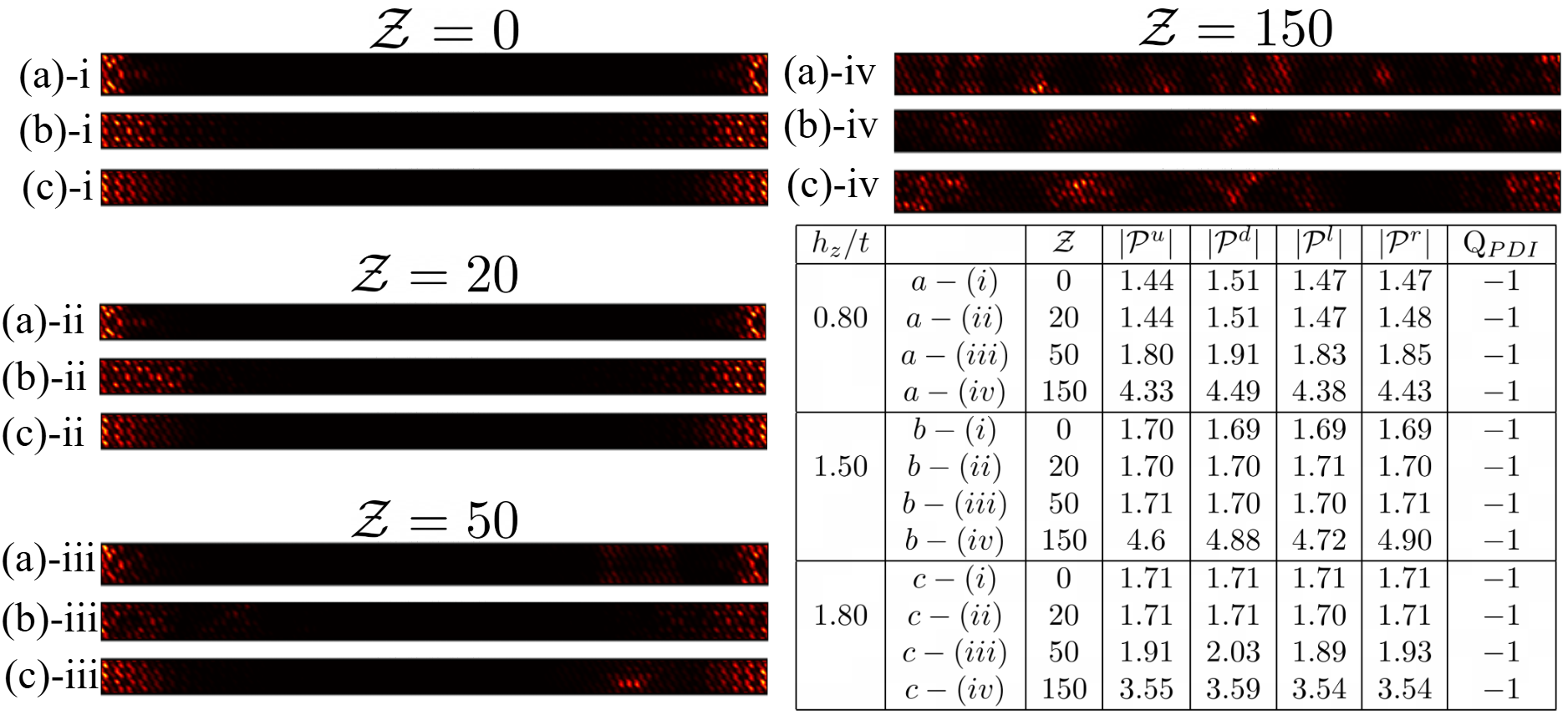}
 \caption{Real-space probability distribution of the low-energy Majorana modes on a \textbf{zigzag} graphene nanoribbon under varying \textbf{out-of-plane magnetic field} and disorder strength. Panels (a)-(c) correspond to increasing values of Zeeman field $h_z/t = 0.80,\ 1.50,$ and $1.80$, respectively, at fixed chemical potential $\mu/t = 2.5$. Each row (i-iv) shows the effect of increasing disorder strength: (i) Clean (no disorder), (ii) Weak disorder ($\mathcal{Z} = 20,\ V_0 = 1.0$), (iii) Moderate disorder ($\mathcal{Z} = 50,\ V_0 = 1.0$), and (iv) Strong disorder ($\mathcal{Z} = 150,\ V_0 = 1.5$). The adjacent table presents the absolute value of the Majorana polarization $|\mathcal{P}|$ for the lower ($\mathcal{P}^d$), upper ($\mathcal{P}^u$), left ($\mathcal{P}^l$), and right ($\mathcal{P}^r$) halves of the strip for each case. \textcolor{black}{The final column of the adjacent table lists the Pfaffian \textcolor{black}{indicator} sector corresponding to each parameter set.}}
 \label{fig:Z_Z_wf}

\end{figure}

\section{ZigZag with out-of-plane Zeeman field}
We now turn to the zigzag strip in the presence of disorder and an out-of-plane magnetic field, as shown in Fig.~\ref{fig:Z_Z_1}. The low-energy spectra in panels (b)-i–iii, obtained at fixed chemical potential $\mu/t = 2.5$, exhibit three regions with zero-energy states. Among these, two field ranges, $h_{z}/t \approx 0.6$–$1.1$ and $1.4$–$2.1$, retain both an open bulk gap and robust zero-energy modes up to moderate disorder, suggesting the stability of Majorana zero modes. The corresponding polarization plots in panels (c)-i–iii confirm this picture: the Majorana polarization remains nearly stable across these field ranges, even as the disorder strength increases. Further insight is obtained from the wavefunction profiles shown in Fig.~\ref{fig:Z_Z_wf}. In contrast to the armchair strip, where the modes are sharply localized at the ends, the Majorana modes in the zigzag geometry are more spatially extended near the edges. Panels (a)-i–iii reveal robust Majorana zero modes persisting up to moderate disorder at both ends of the strip, while panels (b)-ii–iii display modes with extended localization at the center of the strip. At strong disorder, however, the zero-energy states lose their topological protection and evolve into partially separated Andreev bound states, as illustrated in panels (a,b,c)-iv. The phase diagram for the Z-type geometry shows a similar triangular pattern, which becomes increasingly distorted with disorder. Here too, gapped regions host zero-energy states under an out-of-plane magnetic field. However, a clear distinction arises between the in-plane ($y$-direction) and out-of-plane ($z$-direction) field cases: while in-plane fields produce multiple narrow topological transitions that are fragile under disorder, the $z$-field results in broader topological regions, comparable to those in the A-type strip. The associated wavefunctions are less sharply localized, spreading over the edge regions, indicating that although MBS-like states may exist, their localization is weaker compared to the A-type geometry.

Comparing the two edge terminations, we find that while both armchair and zigzag strips support robust Majorana modes over finite ranges of magnetic field and disorder, their spatial character differs substantially. In the armchair case, the modes are sharply localized at the strip ends, whereas in the zigzag case the modes tend to spread along the edges or partially extend into the bulk. This qualitative distinction highlights the role of edge geometry in shaping the robustness and localization properties of Majorana zero modes in graphene nanoribbons.
\end{widetext}

\bibliography{biblio.bib}
\end{document}